\documentstyle[11pt,aaspp4]{article}

\def\go {\mathrel{\raise.3ex\hbox{$>$}\mkern-14mu\lower0.6ex\hbox{$\sim$}}}
\def\lo {\mathrel{\raise.3ex\hbox{$<$}\mkern-14mu\lower0.6ex\hbox{$\sim$}}}

\begin{document}

\title{Thermal and Dynamical Equilibrium in Two-Component Star Clusters}

\author{Wesley A.~Watters\altaffilmark{1}, Kriten J.~Joshi\altaffilmark{2}, 
and Frederic A.~Rasio\altaffilmark{3,4}}

\affil{Department of Physics, Massachusetts Institute of Technology}

\altaffiltext{1}{Present address: Laboratoire de Physique de la Mati\`ere Condens\'ee,
Ecole Polytechnique, 91128 Palaiseau, France; email: ww@pmc.polytechnique.fr.}
\altaffiltext{2}{6-218M MIT, 77 Massachusetts Ave, Cambridge, MA 02139; 
email: kjoshi@mit.edu.}
\altaffiltext{3}{6-201 MIT, 77 Massachusetts Ave, Cambridge, MA 02139; 
email: rasio@mit.edu.}
\altaffiltext{4}{Alfred P.\ Sloan Research Fellow.}

\begin{abstract}

We present the results of Monte Carlo simulations for the dynamical evolution 
of star clusters containing two stellar populations with individual masses 
$m_1$ and $m_2 > m_1$, and total masses $M_1$ and $M_2 < M_1$.  We use both 
King and Plummer model initial conditions and we perform simulations for a wide
range of individual and total mass ratios, $m_2/m_1$ and $M_2/M_1$.  We ignore 
the effects of binaries, stellar evolution, and the galactic tidal field.  The 
simulations use $N = 10^5$ stars and follow the evolution of the clusters until
core collapse.  We find that the departure from energy equipartition in the core 
follows approximately the theoretical predictions of Spitzer (1969) and 
Lightman \& Fall (1978), and we suggest a more exact condition that is based 
on our results.  We find good agreement with previous results obtained by other
methods regarding several important features of the evolution, including the 
pre-collapse distribution of heavier stars, the time scale on which 
equipartition is approached, and the extent to which core collapse is 
accelerated by a small subpopulation of heavier stars. We briefly discuss
the possible implications of our results for the dynamical evolution of primordial
black holes and neutron stars in globular clusters. 

\end{abstract} 


\section{Introduction}

Remarkable advances have been made over the last three decades in our 
understanding of globular cluster dynamics (see, e.g., Meylan \& Heggie
1997 for a recent review). The simple case of a 
two-component cluster is traditionally regarded as the second level of 
sophistication and therefore a logical challenge for new methods that have 
tackled the single-component case.  Two-component clusters were originally 
examined because they better resemble real clusters, which contain a continuous
spectrum of masses.  While somewhat more realistic in this regard, clusters 
with only two mass components still represent a simplification with respect to 
real clusters.  It has been suggested recently, however, that for a range of 
configurations in mass types and the relative size of the two populations, 
two-component clusters can resemble real clusters that are mostly comprised 
of compact objects and main-sequence stars (Kim, Lee, \& Goodman 1998).  
Similarly, clusters containing both single main-sequence stars and
primordial binaries can be modeled, in first approximation, as two-component
systems, although dynamical interactions involving binaries are expected
to play an important role for these systems (Gao et al.\ 1991).
Perhaps the 
best reason to examine a simplified model of any stellar system, however, is 
to obtain a more profound understanding of individual physical processes. 

Much of the discourse regarding two-component systems has focused upon the 
following questions.  First, for what configurations of the cluster is 
dynamical equilibrium precluded (i.e., the system is not stable on dynamical 
time scales)?  Second, for what configurations of the cluster is thermal 
equilibrium precluded (i.e., equipartition of kinetic energies between each 
component is not allowed)?  Both questions originate from an analysis by 
Spitzer (1969), in which he noticed that simultaneous thermal and dynamical 
equilibrium is impossible for some clusters.  In particular, the heavier stars
sink into the center as they lose kinetic energy to the lighter stars during 
the approach to equipartition.  If equipartition is not attained, then the 
heavier stars will continue sinking until their self-gravity dominates the 
cluster's potential in the core.  Shortly thereafter, the heavier component 
will undergo a gravothermal collapse, forming a small dense core comprised 
mainly of the heavier stars (\cite{spitz69}).  Refinements to this analysis 
have obtained similar constraints upon the configurations of two-component 
clusters in dynamical and thermal equilibrium (\cite{lightfall78}).

Several methods have been used to address questions about dynamical and thermal
equilibrium in two-component systems.  These include the construction and study
of one-parameter families of models in dynamical equilibrium (\cite{kondroz82};
\cite{katztaff83}), Monte Carlo approaches to the numerical integration of the 
Fokker-Planck equation (\cite{spithart71}), direct integration of the 
Fokker-Planck equation in phase space (\cite{inagwiy84}; \cite{kimleegood98}), 
and also direct $N$-body simulations (Portegies Zwart \& McMillan 2000).  
The majority of 
work using any one of these methods has been undertaken at least partly in 
order to confirm Spitzer's conclusion (\cite{yosh78}) or refute it 
(\cite{merr81}).  

Dynamical equilibrium is attained and maintained on time scales that are very 
short compared to the amount of time needed for relaxation or equipartition.  
A so-called ``equilibrium model'' (i.e., whose phase-space distribution function 
satisfies the equation for hydrostatic equilibrium) therefore resembles a 
possible stage or snapshot in the evolution of a dynamically stable cluster.  
It is interesting to construct a parametrized family of equilibrium models, 
for which equipartition is either assumed or the temperature ratio allowed to 
vary, in order to determine under what conditions the dynamical equilibrium 
becomes impossible.  In the majority of previous work, the  
distribution functions of such families take the form of lowered Maxwellians 
or spatially-truncated isothermal spheres.
 
Yoshizawa et al.\ (1978) examined a linear series of equilibrium models of 
two-component isothermal spheres with reflecting walls and found that turning 
points of the total energy (stability limits) were positive for some clusters. 
In such cases, a cluster was not self-bounded and would dissociate if the walls
were removed.  These configurations, which were interpreted to preclude 
dynamical equilibrium, occurred under conditions corresponding closely to the 
ones Spitzer proposed.  Katz \& Taff (1983) examined the limits of stability 
for a linear series of equilibrium models of self-bounded two-component 
clusters with a lowered-Maxwellian velocity distribution.  They found that the 
number of possible configurations for clusters in dynamical and thermal 
equilibrium were diminished dramatically in the regime that Spitzer proposed.  
Kondrat'ev \& Ozernoy (1982) constructed a family of equilibrium models, based 
upon a generalization of the single-component King models, 
for which equipartition of kinetic energies was generally impossible.  By 
contrast, Merritt (1981) described a one-parameter family of equilibrium models
for which equipartition in the core is possible for all cluster configurations 
within Spitzer's unstable regime.  It seems clear, however, that these models 
violate an important assumption of Spitzer's analysis, and that they are highly
unrealistic (\cite{merr81}).
   
Few studies that examine the {\it evolution\/} of two-component systems have been
undertaken for a wide range of cluster configurations and using the other
methods mentioned above.  Among the more notable efforts are the Monte Carlo 
calculations of Spitzer \& Hart (1971; later extended to three-component
systems by Spitzer \& Shull 1975), which were interpreted to partially 
confirm Spitzer's original analysis, and the direct Fokker-Planck integrations
of Inagaki \& Wiyanto (1984; see also Inagaki 1985), 
which also partially support Spitzer's claim, 
although by examining a limited range of clusters that do not totally satisfy 
his assumptions.  The former study finds that a single model, which belongs to 
the unstable regime in which equipartition is precluded, develops a collapsing 
subsystem of heavier stars.  Both studies find that central equipartition of 
kinetic energies is never attained throughout the evolution of some clusters.

In this paper we present the results of calculations used to model the 
evolution of two-component clusters with a wide range of configurations.  We
examine several features of the evolution, including the tendency toward 
equipartition, the evolution of mass densities in the core, and the tendency 
for core collapse to be accelerated by the presence of a small subpopulation of
heavier stars.  Our aim is partly to assess the accuracy of Spitzer's analysis 
and also the refinement supplied by Lightman \& Fall (1978).  We have used a 
new Monte Carlo code for modeling the evolution as a sequence of 
equilibrium models whose velocities are perturbed according to the average 
effect of long-range stellar encounters (Joshi, Rasio, \& Portegies Zwart 2000a).  
Our initial two-component systems are isolated King and Plummer models 
with a subpopulation of heavier stars.  

Our paper is organized as follows.  In \S\ref{explanation} we review the 
theoretical arguments that suggest conditions under which simultaneous 
dynamical and thermal equilibrium are not allowed.  In \S\ref{method} we 
discuss our numerical method and the quantities we calculate, and in 
\S\ref{results} we present our main results.  In the last section, 
\S\ref{discussion}, we discuss some additional comparisons between our findings
and those of other studies, as well as some astrophysical implications
of our results for subpopulations of compact objects in globular clusters. 

\section{Equipartition of Kinetic Energies in Two-Component Clusters}
\label{explanation}

According to Spitzer (1969), simultaneous dynamical and thermal equilibrium is 
impossible for two-component star clusters within a certain range of 
configurations.  Let us consider a cluster with stars of two masses, $m_1$ and 
$m_2$, where $m_2 > m_1$.  Moreover, let $M_2$ and $M_1$ be the total mass in 
each component.  Spitzer assumed that $M_2 \ll M_1$, which is normally the case
for real clusters.  He concluded that for certain 
values of the ratios $M_2/M_1$ and $m_2/m_1$, equipartition will not be 
attained as the heavier and lighter stars exchange kinetic energy, and hence 
the heavier stars will sink very far into the center.  Moreover, the heat 
exchange with lighter stars promotes them to higher orbits, so that eventually 
insufficient numbers will remain in the center to conduct heat rapidly away 
from the heavier stars.  If the mass-stratification proceeds far enough,
then the self-gravity of the heavier stars will dominate the potential in the 
core, and this subsystem will undergo gravothermal collapse.  The result is
a very dense core comprised exclusively of heavier stars.  

In his analysis, Spitzer begins by assuming global equipartition.  Global
equipartition is not realistic, however, because the relaxation and 
equipartition times vary greatly throughout the cluster, becoming
longer than the age of the universe in the outer halo.  In fact, we expect 
equipartition only in the inner region where relaxation times are shortest.  
His discussion is mostly unchanged by this, so long as we confine its relevance
to processes in this inner region.  We shall reproduce here only the main 
strategy of his argument and its conclusions.  Let $v_{m1}^2$ and $v_{m2}^2$ 
represent the mean-square velocities of stars in each component.  As mentioned,
The assumption of equipartition implies that the temperature ratio $\xi$ is 
equal to unity,
\begin{equation}\label{equip}
\xi \equiv \frac{(1/2)m_2v_{m2}^2}{(1/2)m_1v_{m1}^2} = 1 \, \rm{.}
\end{equation}
Spitzer then applies a component-wise virial theorem, which for the heavier 
component gives
\begin{equation}\label{virial}
v_{m2}^2 = \frac{(2/5)GM_2}{r_2} + \frac{G}{M_2}\int_0^\infty\frac{\rho_2M_1(r)dV}{r}, 
\end{equation}     
where the first term represents the gravitational self-binding energy of the 
heavier stars, $r_2$ is the virial radius of the heavier stars, 
$\rho_2$ is the density of the heavier stars at a distance $r$ 
from the center, $M_1(r)$ is the total mass of the heavier stars contained 
within the radius $r$, and $dV$ is the volume element.  Spitzer assumes that,
since $M_2 \ll M_1$, the second term in the corresponding equation for 
$v_{m1}^2$ can be ignored, and also that, since $m_2 \gg m_1$, the heavier
stars will be concentrated in the center of the system and $M_1(r)$ may be 
approximated by $4\pi\rho_1(0)r^3/3$ in equation~(\ref{virial}).  Spitzer \& 
Hart (1971) noticed that the second assumption often does not hold strictly.  
In particular, even far into the evolution many heavier stars can still reside 
well outside of the core.  Merritt (1981) constructed equilibrium models that 
violate this assumption by great amounts and discovered that equipartition is 
possible for some (admittedly unrealistic) configurations which can be realized 
for all values of $m_2/m_1$ and $M_2/M_1$. 

Under Spitzer's assumptions, after a series of manipulations one obtains the 
following expression for the quantity $S$, as a direct consequence of 
equation~(\ref{equip}),
\begin{equation}\label{pent}
S \equiv \left(\frac{M_2}{M_1}\right)\left(\frac{m_2}{m_1}\right)^{3/2} 
= \frac{(\rho_{h1}/\rho_{h2})^{1/2}}{(1 + \alpha\rho_{h1}/\rho_{h2})^{3/2}},
\end{equation}
where $\rho_{h1}$ and $\rho_{h2}$ are the densities within the half-mass radii 
for each component, and where $\alpha$ depends on the distribution of mass 
within the cluster.  In particular, if we denote by $r_{m2}$ and $r_{h2}$ the 
rms and half-mass radius for the heavier component, respectively, then $\alpha$
is given by
\begin{equation}
\alpha = \frac{5\rho_1(0)}{4\rho_{h1}}\left(\frac{r_{m2}}{r_{h2}}\right)^2.
\end{equation}
Spitzer estimates a value of 5.6 for $\alpha$.  Merritt (1981) contests the 
value 3.5 assigned by Spitzer to $\rho_1(0)/\rho_{h1}$ for polytropes with 
$n$ between 3 and 5, but finds that it corresponds approximately to its 
minimum value, which, as we shall see, does not change Spitzer's conclusion.  
In particular, since the RHS of equation~(\ref{pent}) has a maximum value 
$0.38\alpha^{-1/2}$ with respect to variation in $\rho_{h1}/\rho_{h2}$, it 
follows that equipartition is possible only if $S$ does not exceed a 
critical value $\beta \equiv 0.38\alpha^{-1/2}$,
\begin{equation}\label{critical}
S < \beta, \mbox{  where } \beta = 0.16 \mbox{ for }\alpha = 5.6.
\end{equation}
Spitzer suggests that for smaller values of the individual mass ratio 
$m_2/m_1$, the inequality~(\ref{critical}) remains valid, except that $\beta 
\rightarrow 1$ as $m_2/m_1 \rightarrow 1$.  

While it is useful to assess under what conditions simultaneous dynamical and 
thermal equilibrium are not expected, it is also interesting for our purposes to 
estimate the extent of departure from thermal equilibrium where dynamical 
equilibrium holds.  To that end, let us maintain the condition stated in 
(\ref{virial}), while not insisting that $\xi = 1$ in equation~(\ref{equip}).  
By following the value $\xi$ through Spitzer's analysis, we find that the
temperature ratio has a lower bound for a given value of $S$,
\begin{equation}\label{lowtrat}
\xi > \left(\frac{S}{\beta}\right)^{2/3}.
\end{equation}
    
Inagaki \& Wiyanto (1984) performed an analysis similar to Spitzer's, except by
casting equations~(\ref{equip}) and~({\ref{virial}) in terms of core values for
the component-wise total mass, mean-square velocity, and density.  They 
obtained a minimum difference between the core temperatures of each component. 
Letting $M_{c2}$ and $M_{c1}$ denote the total mass contained within the core 
for the heavier and lighter components, respectively, and letting $r_{c1}$ 
represent the core-radius of the lighter stars, this difference is given by:  
\begin{equation}\label{inagdiff}
\frac{1}{2}m_2(v_{m2}^2)_{{\rm min}} - \frac{1}{2}m_1v_{m2}^2 = \left[\frac{1}{2}\left(\frac{M_{c2}}{M_{c1}}\right)^{2/3} - 0.2m_1\right]\frac{GM_{c1}}{r_{c1}}.
\end{equation} 
For now, we note that the minimum ratio (given by the inequality~[6]), 
and difference (given by eq.~[7]) of component-wise temperatures increases 
with increasing values of the total mass 
ratio, and that this increase is steeper for increasing $m_2/m_1$ 
(in eq.~[6], recall that $\beta \rightarrow 1$ as $m_2/m_1 \rightarrow 1$).

Lightman \& Fall (1978) developed an approximate theory for the core collapse 
of two-component clusters which resembles that of Spitzer.  They examined two 
constant-density isothermal spheres representing the cores of the heavier 
and lighter components, where the radius of the former is smallest.  By 
applying a component-wise virial theorem and several simplifying assumptions, 
they find a set of four ordinary differential equations for the virial radii 
and total masses in each component.  They obtain the following condition for 
equipartition of energies between the two components, where we let $\tilde{m} 
\equiv m_2/m_1$ and $\tilde{M} \equiv M_2/M_1$:
\begin{equation}\label{Gamma}
\Gamma(\tilde{m},\tilde{M}) \equiv \tilde{m}^3\tilde{M}^2
\left[\frac{27}{4}\left(1 + \frac{3\tilde{M}}{2\tilde{m}}\right)
\left(1 + \frac{5}{2}\tilde{M}\right)^{-3}\right] \leq 1 \, \rm{.}
\end{equation}
Solutions to the differential equations exhibit a minimum temperature ratio 
$\xi_{{\rm min}},$ which they suggest bears the following approximate relation 
to $\Gamma$,
\begin{equation}\label{Gamin}
\xi_{{\rm min}} \simeq \Gamma^{1/3}. 
\end{equation}
In this case also, the minimum temperature ratio increases with increasing 
values of the individual and total mass ratios.  

The majority of attempts to 
evaluate the theoretical predictions of Spitzer (1969) and Lightman \& 
Fall (1978) have met with moderate success.  For example, Yoshizawa 
et al.\ (1978) obtained 0.25 for the value of $\beta$ in the case of 
spatially-truncated two-component isothermal spheres.  Nevertheless, few 
investigations have examined models with a comprehensive range of individual 
and total mass ratios (outside of studies based upon turning points along a 
sequence of equilibrium models).  In the next two sections we discuss the 
methods we used and results that we obtained for a relatively broad survey of 
the parameter space determined by $M_2/M_1$ and $m_2/m_1$.  
   
\section{Numerical Methods and Definitions}\label{method}

We used a Monte Carlo method for modeling the dynamical evolution of clusters
as a sequence of equilibrium models subject to regular velocity perturbations.
The velocity perturbations represent the average effect of long-range stellar 
encounters (\cite{hen71}). Our Monte Carlo code has been described in detail by 
Joshi, Rasio, \& Portegies Zwart (2000a) and Joshi, Nave, \& Rasio (2000b).  
The code allows us to perform dynamical simulations for realistic clusters
containing up to $N \sim 10^5 - 10^6$ stars on a parallel supercomputer.
In the present study, 
we ignore the effects of binaries, stellar evolution, and the galactic 
tidal field.  
Our units are defined in Joshi et al.\ (2000a, \S 2.8).
The unit of length is close to the virial radius of the cluster, the mass
is measured in units of the total initial cluster mass, and the unit
of time is of order the initial half-mass relaxation time $t_{\rm rh}$.
In this paper, when reporting times in units of $t_{\rm rh}$, we have adopted
the same definition used in previous studies of
two-component clusters since Spitzer \& Hart (1971), 
\begin{equation}
t_{\rm rh} = \frac{0.06M^{1/2}r_h^{3/2}}{G^{1/2}m\log_{10}(0.4N)},
\end{equation}
where $M=M_1+M_2$ is the total cluster mass, $r_h$ is the initial
cluster half-mass radius, and $m=M/N$ is the average stellar mass.

We undertook two sets of calculations, hereafter called sets $A$ 
and $B$. All calculations are performed for a cluster containing $N = 10^5$ 
single stars. The initial model used in each calculation of set $A$ was a 
two-component King model (\cite{king66}). In particular, the velocities and
positions for all stars with a mass $m_1$ were chosen according to the 
King model distribution function with dimensionless central potential 
$W_0 = 6$. Although the initial King model is truncated at its finite 
tidal radius, we do not enforce that tidal boundary during the evolution, 
allowing the cluster to expand indefinitely. 
In each case, some fraction of the stars were 
then changed to a mass $m_2 > m_1$ according to a chosen value of the total 
mass ratio $M_2/M_1$.  The initial ratio of mean temperatures in the heavier 
and lighter components was therefore $m_2/m_1$.  The initial models for 
calculations in the set $B$ were generated in a similar way, except using the 
Plummer distribution function (polytrope with $n=5$; see, e.g., Binney \&
Tremaine 1989).  Both sets of 
calculations are listed in Table~\ref{tab_results}.  
The set $A$ includes calculations undertaken for a range of total 
mass ratios ($M_2/M_1$ = $3 \times 10^{-3}$ to $0.6$) and a range of individual mass 
ratios ($m_2/m_1$ = 1.5 to 6).  The set $B$ comprises only 9 systems, including 
4 studied by Inagaki \& Wiyanto (1984), all with $m_2/m_1 = 2$.  
Every calculation is terminated at core collapse, measured at the instant that 
a number density of $10^8$ (in our units) is attained in the core. Our results
are not sensitive to the exact value of the core density used to terminate
the calculation. However, the value of the core-collapse time $t_{\rm cc}$
determined numerically can have a large statistical uncertainty, particularly when the
core is dominated by a small number of heavier stars near the end of the
evolution (in those cases we estimate that the statistical uncertainty on the values
of $t_{\rm cc}/t_{\rm rh}$ reported in Table~1 can be as large as $\sim5$\%).
The majority of our calculations require between $10-20\,$CPU hours
on an SGI/Cray Origin2000 supercomputer to reach core collapse.   

The range of values we consider for $m_2/m_1$ and $M_2/M_1$ includes a number of
astrophysically relevant cases. In particular, a subpopulation of neutron
stars in a dense globular cluster might have $m_2/m_1 \simeq 2$ 
(e.g., corresponding to $m_2=1.4\,M_\odot$ for an average background stellar
mass $m_1=0.7\,M_\odot$) and 
$M_2/M_1 \sim 10^{-3} - 10^{-2}$ depending on the neutron star retention
fraction. A subpopulation of black holes might have $m_2/m_1 \simeq 5-10$
and $M_2/M_1 \sim 10^{-3} - 10^{-2}$. Massive blue stragglers
or primordial binaries containing main-sequence stars near the turn-off mass,
would have $m_2/m_1 \simeq 2-3$ and $M_2/M_1 \sim 10^{-3} - 10^{-1}$.

For each calculation we record several quantities at each program time step 
(the time steps are proportional to a fraction of the core relaxation time;
see Joshi et al.\ 2000a).  
These continuous measurements include the total-cluster core radius and several
component-wise Lagrange radii.  Within each of these radii we also count the 
number of stars and calculate the mean temperature and mean mass density for
each component.  Of particular interest are the quantities calculated inside 
the core radius, where relaxation times are shortest and where thermal 
equilibrium is the most likely to occur. We use the customary definition 
for the total-cluster core radius $r_c$ (\cite{spitz87}),
\begin{equation}
r_c = \left(\frac{3v_m(0)^2}{4\pi G\rho(0)}\right)^{1/2},
\end{equation}
where $v_m(0)^2$ is the mean-square velocity and $\rho(0)$ the mean density of
stars at the center.  We calculate the ratio of core densities in each 
component $(\rho_{c2}/\rho_{c1})$, and also the ratio of core temperatures.  
We also calculate the minimum core temperature ratio $\xi_{{\rm min}}$ that is
reached after approximately 90\% of the pre-collapse evolution in all cases.  
Specifically, $\xi_{{\rm min}}$ is calculated as the average temperature ratio
from 90\% to 95\% of the core collapse time.  The core collapse time 
$t_{{\rm cc}}$ and the time $t_{{\rm \rho}}$ at which the core mass densities 
of each component become equal (i.e., $\rho_{c2}/\rho_{c1} = 1$) are also 
measured.  We report our main results in Table~\ref{tab_results}, and 
we discuss these in the following section.

\section{Results}\label{results}     

The evolution of the core temperature ratio $\xi$ is shown in 
Figure~\ref{3temprats} for three calculations in the set $A$ (two-component King 
models): namely, for $S = 0.05$ (top), $S = 0.5$ (middle), and $S = 1.24$ 
(bottom).  Figure~2 shows the core temperatures of the 
lighter stars (top) and the heavier stars (bottom) for the case $S = 1.24$.
Several features that we expect and that have been mentioned already in
\S\ref{explanation} are easily recognizable.  The temperature ratio begins 
with a value $m_2/m_1$ and then decreases gradually as equipartition is 
approached.  It is clear that $\xi$ reaches a minimum value that is greater 
than 1 for the case $S = 1.24$, so that equipartition is clearly never attained. 
Equipartition is nearly attained for $S = 0.5$, and $\xi_{{\rm min}} = 1$ to 
within 5\% for $S = 0.05$.  It is clear from Figure~\ref{unstable_temps} that 
the heavier component cools initially, then maintains a constant mean kinetic 
energy, and then begins to heat prior to core collapse.  At the same time, the 
lighter component steadily becomes hotter as it receives energy from the 
heavier component.  The temperature ratio in the last time steps becomes very 
noisy because the temperatures are computed using the relatively few stars that
remain in the core.   

The temperature ratio $\xi$ reaches a minimum value at different times with 
respect to core collapse for each of the models shown in Figure~\ref{3temprats}.  
In cases where the minimum value is greater than 1, it 
sometimes appears that the gravothermal catastrophe has beaten the approach to 
equipartition.  In such cases, one may ask whether an initial model with a less
concentrated spatial distribution and a different initial relaxation time would
yield a different minimum temperature ratio.  In fact, we find that 
$\xi_{{\rm min}}$ is robust with respect to changes in the initial value of the
dimensionless potential $W_0$.  The evolution of the temperature ratio for 
three calculations, where $W_0 = 1$, 5, and 10 for $S = 1$ and $m_2/m_1 = 5$, 
are shown in Figure~\ref{3temprats_3Ws}.  In all three cases the minimum 
temperature ratio is approximately 1.55.    

The evolution of the core mass density ratio $\rho_{c2}/\rho_{c1}$ is shown 
in Figure~\ref{3densrats} for the same three clusters and in the same order.  
The core mass densities of each component are shown in Figure~\ref{unstable_dens} 
for the model with $S = 1.24$.  One can see clearly that, as $S$ increases,
the core mass densities in each component become equal sooner with respect to
core collapse.  That is, for larger $S$ the self-gravity of the heavier stars 
dominates the potential in the core for a longer period prior to the onset of 
the gravothermal catastrophe.  Moreover, we can see that, although the core 
density is initially dominated by the lighter stars, the 
heavier stars overtake and exceed the density of lighter stars by more 
than an order of magnitude prior to core collapse for $S = 0.5$ and $S = 1.24$.
     
Approximate values of the minimum core temperature ratio $\xi_{{\rm min}}$ 
are plotted using three symbols in the parameter space determined by $M_2/M_1$ 
and $m_2/m_1$ in Figure~\ref{temps} for 37 calculations in the set $A$.  Also 
drawn are the Spitzer and Lightman-Fall ``stability boundaries,'' above which 
simultaneous dynamical and thermal equilibrium are supposedly prohibited
($S = 0.16$ and $\Gamma = 1$, respectively; cf. eqs.~[5] \& [8]).  
Our simulations allow us to determine $\xi_{{\rm min}}$ 
with a numerical accuracy of about 5\%. Specifically, $\xi_{{\rm min}}$ is calculated as 
the average core temperature ratio from 90\% to 95\% of the pre-collapse 
evolution, and this average has a standard deviation of approximately 0.05
in our calculations for $N = 10^5$ stars.
Therefore, those calculations marked with a ``$\Box$'' in Figure~\ref{temps}
have been determined to reach equipartition within our numerical accuracy. 
One can see that the Spitzer boundary $S = 0.16$ is approximately respected for 
$m_2/m_1 \geq 2$.  By comparison, the Lightman-Fall boundary falls well inside 
the range of clusters which have clearly not attained equipartition.  In spite 
of this, the Lightman-Fall boundary better reproduces the shape of boundaries 
between regions of constant $\xi_{{\rm min}}$.  A more properly-drawn Spitzer 
boundary has a similar shape, recalling that $\beta \rightarrow 1$ as $m_2/m_1 
\rightarrow 1$.  Based upon the results shown in Figure~\ref{temps}, we 
propose our own condition for equipartition, 
\begin{equation}\label{Lambda}
\Lambda \equiv \left(\frac{M_2}{M_1}\right)\left(\frac{m_2}{m_1}\right)^{2.4} 
\geq 0.32 \rm{.}
\end{equation}
The boundary determined by equation~(\ref{Lambda}) is strictly valid for 
$1.75 < m_2/m_1 < 7$, and is also drawn in Figure~\ref{temps}. For 
$ m_2/m_1 < 1.75$, equipartition is achieved for all clusters considered.
For $ m_2/m_1 > 1.75$, extrapolation of equation~(\ref{Lambda}) seems
reasonable.

The dependence of $\xi_{{\rm min}}$ on $S$ for several values of $m_2/m_1$
is shown in Figure~\ref{mintemps_S}.  Also drawn is the Spitzer stability 
boundary.  These curves are broadly consistent with trends anticipated by the
inequality~(6).  In particular, $\xi_{{\rm min}}$ increases 
with $S$, and the initial slope of this increase becomes larger with increasing
$m_2/m_1$ (again recalling that $\beta \rightarrow 1$ as $m_2/m_1 \rightarrow 
1$). The dependence of $\xi_{{\rm min}}$ on $\Gamma$ for several values 
of $m_2/m_1$ is shown in Figure~\ref{mintemps_Gamma}.  Also drawn is the value
of $\xi_{{\rm min}}$ anticipated by equation~(\ref{Gamin}), and the 
Lightman-Fall stability boundary.  One can see that, while the numerical
results are displaced from the predicted curve, they have the same curvature 
and display a similar trend.  The minimum temperature {\it difference\/} 
for calculations in 
the set $B$ (two-component Plummer models) are shown in Figure~\ref{compare_plot}, 
where they are compared with the results of Inagaki \& 
Wiyanto (1984).  Here, agreement is excellent except in the limit of small 
$M_2/M_1$, where the temperature of the heavier component is 
calculated using very few stars, so that the difference is characterized 
by a large amount of noise. As a last comparison, we note that, using a Monte 
Carlo scheme different from 
ours, Spitzer \& Hart (1971) found that $\xi_{{\rm min}} = 1.34$ for a Plummer
model with $S=1.24$ and $m_2/m_1 = 5$, whereas for the same system we obtain 
$\xi_{{\rm min}} = 1.60$.      

Recall that $t_{{\rm \rho}}$ is the time at which the mass densities of each 
component become equal in the core.  Approximate values of the time 
$t_{{\rm \rho}}$ as a fraction of the core collapse time $t_{{\rm cc}}$ are 
plotted using three symbols in the parameter space determined by $M_2/M_1$ and 
$m_2/m_1$ in Figure~\ref{density_tcc}, for 37 calculations in the set $A$.  
This plot confirms the trend anticipated by the previous examination of three 
individual clusters (Fig.~4).  Namely, the amount of time --- as a fraction of the 
core-collapse time --- during which the heavier stars dominate the potential in the 
core increases with $S$.  The trends appear not to respect any of the 
previous stability
boundaries very well, but our condition~(12) fares best.  
They may nevertheless shed light on the related question
of whether a dense subsystem of heavy stars collapses {\it independently\/}, 
since the self-gravity of the heavier stars must dominate the potential in the 
core in order for this to happen.  Where equipartition is attained, the 
mass segregation is retarded or stopped, so that equal mass densities may not 
occur until core collapse (i.e., $t_{{\rm \rho}} \simeq t_{{\rm cc}}$).      

All of the calculations were terminated at core-collapse, at which time the 
radius containing 1\% of the mass in the heavier component diminishes sharply. 
The time $t_{{\rm cc}}$ is measured at the instant when a number density of 
$10^8$ in our units is attained within the core (see \S 3). The variation of the core 
collapse time $t_{{\rm cc}}$ with $S$ for several values of $m_2/m_1$ is shown in 
Figure~\ref{tcc_times}.  The trends confirm that the onset of core collapse 
is accelerated by the presence of a small and heavier subpopulation, in 
agreement with the findings of others (\cite{inagwiy84};
Inagaki 1985; \cite{quin96}).

\section{Discussion}\label{discussion}

In this section we discuss several features of the evolution in more detail,
and we mention some possible astrophysical applications of our results to
the evolution of compact stellar remnants in globular clusters.

The temperature ratio $\xi$ in Figure~\ref{3temprats} initially has the value 
$m_2/m_1$.  While this is an artifact of the way our initial models were 
constructed, $m_2/m_1$ happens also to be the most realistic value of $\xi$ for
equilibrium models with a relatively shallow potential.  In families of 
equilibrium models it is typical to find that $\xi \rightarrow m_2/m_1$ as 
$W_0 \rightarrow 0$ (\cite{kondroz82}, \cite{katztaff83}).  In trials for which
initial models were modified so that $\xi$ had some value other than $m_2/m_1$,
a brief period of rapid relaxation ensued which restored the value $m_2/m_1$.
This effect has been observed in simpler models of the evolution calculated 
using other methods (\cite{lightfall78}).  

In the core, the initial behavior of the temperature ratio is mostly 
determined by the temperature of the heavier component, while the mean kinetic 
energy of the lighter stars, which are more abundant at first, increases 
gradually (Fig.~\ref{unstable_temps}).  Spitzer (1969) suggested that the 
approach to equipartition is characterized by the exponential decay of kinetic 
energy in the heavier component, with a time constant equal to {\it twice\/} the 
so-called equipartition time,
\begin{equation}\label{teq}
t_{{\rm eq}} = t_{r1}\frac{3\pi^{1/2}}{16}\frac{m_1}{m_2}
\left(1 + \frac{v_{m2}}{v_{m1}}\right)^{3/2},
\end{equation} 
where $t_{r1}$ is a relaxation time for the stars of mass $m_1$.  In the case 
where the mean-square velocities of each component are initially equal, the 
initial equipartition time is $t_{{\rm eq}} \simeq t_{r1}(m_1/m_2)$.  (It 
should be noted that $t_{{\rm eq}}$ decreases as equipartition is approached.)
This characterization of the decline in kinetic energy of heavier stars agrees
very well with our results for stars contained within the half-mass radius,
where we assume $t_{r1} \simeq t_{{\rm rh}}$, the initial half-mass relaxation
time for the cluster as a whole.  In particular, the kinetic energy of the 
heavier component diminishes to a fraction $1/e$ of its initial value 
(after subtracting its minimum value for the entire evolution) within 
$0.39\,t_{{\rm rh}}$ for $S = 1$, $m_2/m_1 = 5$, and within $1.3\,t_{{\rm rh}}$ 
for $S = 0.5$, $m_2/m_1 = 1.5$, in good agreement with the theory (which 
predicts a time $2t_{{\rm eq}} \simeq 2(m_1/m_2)t_{rh}$).  However, we 
find that equipartition is approached on a similar time scale in the core, 
where the theory predicts that $t_{{\rm eq}}$ should be shorter by 
approximately $1/5$, and hence agreement is poor (the ratio of 
initial core and half-mass relaxation times for King models
with $W_0 = 6$ is approximately 1/5; see \cite{quin96}).  

A leveling in the temperature ratio at a minimum value greater than 1 has been 
observed in calculations using other methods as well (\cite{inagwiy84}, 
\cite{lightfall78}). We find that this leveling is approximately coincident 
with the heavier stars reaching their maximum numbers within the core.
Inagaki \& Wiyanto (1984) found 
that an increase in the core temperature ratio prior to or during collapse is 
coincident with $t_{{\rm \rho}}$, the time at which equal mass densities are 
attained in the core.  While we are not able to resolve adequately the 
late-collapse behavior of $\xi$ (because our calculation loses accuracy in this
regime), it is clear from Figure~\ref{unstable_temps} that the heavier 
component begins to heat prior to collapse.  Indeed, we find that the 
temperature of the heavier component does not begin to rise until 
$t > t_{{\rm \rho}}$.

Katz \& Taff (1983) examined the turning points in a linear series of 
equilibrium models.  In particular, they studied a one-parameter family of 
self-bounded isolated equilibrium models with a lowered-Maxwellian velocity 
distribution.  Turning points representing the limits of stability for 
dynamical equilibrium were obtained for several values of $m_2/m_1$ and 
$M_2/M_1$ in terms of maximum possible values of the dimensionless potential 
$k$.  Katz \& Taff also calculated the core temperature ratio $\xi$ that 
corresponds to each maximum value of $k$.  Since for their models $\xi$ was 
found to approach 1 for large values of $k$, the core temperature ratios 
calculated for each turning point represent the minimum allowed value of $\xi$ 
for given values of $m_2/m_1$ and $M_2/M_1$.  These are plotted with respect 
to $S$ for several values of $m_2/m_1$ in Figure~\ref{mintemps_S_katztaff} and 
should be compared with our results, shown in Figure {\ref{mintemps_S}}.  While
these numbers exhibit a similar trend with $S$ for given $m_2/m_1$, the trend 
in $m_2/m_1$ evidently disagrees with our results, and therefore also the 
prediction of the inequality~(\ref{lowtrat}).  It is important to bear in mind,
however, that these ratios are obtained just for the members of this particular
family, which do not necessarily represent states that can be obtained by real 
dynamical processes.  Our results contradict those of Katz \& Taff (1983) 
insofar as we find clusters with smaller values of $\xi_{{\rm min}}$ for 
identical values of $m_2/m_1$ and $M_2/M_1$, and which appear to be stable on 
dynamical time scales.  Katz \& Taff found that the number of possible 
configurations for their models in dynamical equilibrium diminish sharply for 
$S > 0.16$.  It is interesting to note also that the values they obtain for 
$\xi_{{\rm min}}$ appear not to diminish below 1.10 for small values of 
$M_2/M_1$.

We concur with the findings of Spitzer \& Hart (1971) that many heavier stars 
remain outside the core throughout the evolution.  This is clear from plots of 
Lagrange radii as a function of time for the heavier component
(Fig.~\ref{S1.241u5extra_lagrange}).  This casts doubt on the assumption,
committed in Spitzer's original analysis, that all of the heavier stars quickly
become concentrated in the core (see \S\ref{explanation}).  In particular, we 
find that for $S = 1.24$, $M_2/M_1 = 0.111$, and $m_2/m_1 = 5$, the radius 
containing $75\%$ of the mass in the heavier component diminishes to only 
$50\%$ of its initial value (and hence remains larger than the core radius) 
throughout the evolution.  Nevertheless, by the onset of core collapse, we 
frequently observe for calculations with large $m_2/m_1$ that no 
lighter stars remain in the core. 

Our results may have important implications for the dynamical evolution
of various subpopulations of interesting objects in globular clusters.
In particular, a subcomponent of primordial black holes with 
$m_2/m_1\simeq10$ is expected to
remain far from energy equipartition with the rest of the cluster, and to evolve 
very quickly to core collapse on its own relaxation time scale.
For a typical cluster IMF (initial mass function), and assuming that all
black holes formed initially by the stellar population are retained in the cluster,
we expect $M_2/M_1\simeq10^{-3}-10^{-2}$, well above our stability boundary
in Figure~6. Recent $N$-body simulations for clusters containing primordial
black holes indicate that, after reaching core collapse, the dense subcluster
of black holes evaporates quickly in the background cluster potential.
Three-body processes occurring in the post-collapse phase produce  
a significant number of tight black hole binaries that will coalesce in 
a few billion years, making these binaries important sources of gravitational
waves for current ground-based laser interferometers (Portegies Zwart \&
McMillan 2000; see also Kulkarni, Hut, \& McMillan 1993 and Sigurdsson \&
Hernquist 1993). For neutron stars, with $m_2/m_1\simeq2$, our results
suggest that equipartition can be reached if the fraction of the total
cluster mass in neutron stars is $\lo5\%$. This fraction is higher than
would be predicted for a standard cluster IMF and neutron star progenitor
masses (even if, in contrast to what is suggested by many observations,
neutron stars were born without the kicks that might eject a large fraction
from the cluster). However, many multi-mass King models and dynamically
evolving Fokker-Planck models of globular clusters based on fits to both
photometric and kinematic data suggest that much higher fractions of
neutron stars may be present in many clusters. 
For example, the recent Fokker-Planck models of Murphy et al.\ (1998) for
47~Tuc suggest that 4.6\% of the total cluster mass is in the form of
dark stellar remnants of mass $1.4\,M_\odot$. With such a high mass
fraction, it is possible that the neutron stars in 47~Tuc may remain
out of energy equipartition with the rest of the cluster.
More sophisticated dynamical simulations taking into account the full mass
spectrum of the cluster, stellar evolution, and binaries, will be necessary
to resolve the issue.

\section{Summary and Conclusions}\label{conclusions}

Although we have omitted the effects of binaries (which have been shown to 
retard the onset of core collapse, see, e.g., Gao et al.\ 1991) and also stellar 
evolution (which can have important implications for the early evolution, see 
Joshi et al.\ 2000b), the following observations and conclusions seem 
justified:

\begin{itemize}

\item
For some two-component clusters the core temperature ratio becomes constant 
over some fraction of the evolution at a minimum value greater than 1, in 
agreement with previous results obtained using other methods.  

\item
The departure from equipartition calculated for a range of individual and total 
mass ratios approximately respects the theoretical predictions of Spitzer 
(1969) and Lightman \& Fall (1978).  The agreement with Spitzer is reasonable 
for  $m_2/m_1 \geq 2$,  and the Lightman-Fall stability boundary 
($\Gamma = 1$) appears to reflect the shape of regions of constant 
$\xi_{{\rm min}}$ in the parameter space determined by $M_2/M_1$ and $m_2/m_1$,
although it lies well inside the region occupied by clusters for which 
equipartition is clearly not attained.  A more accurate boundary that is 
suggested by our results is given by equation~(\ref{Lambda}).  The trend in the
minimum values predicted for the temperature ratio by Lightman \& Fall (1978)
is similar to what we observe.  

\item
Stars in the heavier component do not 
immediately fall into the cluster core as assumed by Spitzer in his analysis, and 
instead many remain well outside the core throughout the evolution.  

\item 
The approach to equipartition within the half-mass radius appears to occur on the 
time scale $2t_{{\rm eq}}$, as suggested by Spitzer (1969; see eq.~[\ref{teq}]).  

\item
A core temperature ratio of $m_2/m_1$ appears to be a robust
quantity for equilibrium models with a relatively shallow potential, in 
agreement with previous results obtained using other methods.   

\item 
For clusters with $M_2/M_1 \ll 1$ and $m_2/m_1 > 1$, core collapse times decrease 
with increasing $M_2/M_1$ and $m_2/m_1$, in agreement with previous results.

\end{itemize}

\acknowledgments

We are grateful to Steve McMillan for providing us the software 
used in constructing our initial conditions, and to Simon Portegies Zwart for many 
helpful conversations. We also thank John Fregeau for helping us with some 
of the numerical integrations.
This work was supported by NSF Grant AST-9618116 and 
NASA ATP Grant NAG5-8460. F.A.R.\ is supported in part by an Alfred P.\ Sloan 
Research Fellowship.  Computations were performed under grant AST980014N from
the National Computational Science Alliance and utilized the SGI/Cray Origin2000 
supercomputer at Boston University.

\clearpage

\begin{deluxetable}{rrrrrrrc}
\footnotesize
\tablecaption{Models and Results\label{tab_results}}
\tablewidth{0pt}
\tablehead{
\colhead{$S$} & \colhead{$m_2/m_1$} & \colhead{$M_2/M_1$} & 
\colhead{$N_2$} & \colhead{$\xi_{min}$} & \colhead{$t_{{\rm \rho}}/t_{{\rm rh}}$} & 
\colhead{$t_{{\rm cc}}/t_{{\rm rh}}$} & \colhead{Model}
}
\startdata
0.05 &1.50 &0.0272 &1,782 &1.010 &12.8 &12.8 &King\nl
0.05 &1.75 &0.0216 &1,219 &1.020 &12.0 &12.0 &King\nl
0.05 &2.00 &0.0176 &876 &1.028 &11.2 &11.2 &King\nl
0.05 &3.00 &0.00962 &320 &1.027 &9.2 &9.2 &King\nl
0.05 &4.00 &0.00625 &156 &1.016 &8.7 &9.0 &King\nl
0.05 &6.00 &0.00340 &57 &1.018 &8.2 &8.2 &King\nl
0.10 &1.50 &0.0544 &3,502 &1.023 &12.3 &12.3 &King\nl
0.10 &1.75 &.0432 &2409 &1.035 &10.9 &11.1 &King\nl
0.10 &2.00 &0.0354 &1,737 &1.043 &9.2 &9.6 &King\nl
0.10 &3.00 &0.0192 &637 &1.039 &5.9 &6.4 &King\nl
0.10 &4.00 &0.0125 &311 &1.053 &4.9 &5.5 &King\nl
0.10 &6.00 &0.00680 &113 &1.061 &3.7 &4.3 &King\nl
0.15 &1.50 &0.0816 &5,162 &1.027 &11.5 &11.7 &King\nl
0.15 &1.75 &0.0648 &3570 &1.043 &9.7 &10.1 &King\nl
0.15 &2.00 &0.0530 &2,583 &1.044 &7.5 &8.5 &King\nl
0.15 &3.00 &0.0289 &953 &1.082 &4.4 &5.3 &King\nl
0.15 &4.00 &0.0188 &467 &1.094 &3.3 &3.9 &King\nl
0.15 &6.00 &0.0102 &170 &1.128 &2.4 &2.6 &King\nl
0.20 &1.50 &0.109 &6,767 &1.021 &11.0 &11.5 &King\nl
0.20 &1.75 &0.0864 &4,704 &1.049 &8.4 &9.5 &King\nl
0.20 &2.00 &0.0707 &3,415 &1.052 &6.8 &7.8 &King\nl
0.20 &3.00 &0.0385 &1,267 &1.090 &3.8 &4.6 &King\nl
0.20 &4.00 &0.0250 &621 &1.124 &2.6 &3.2 &King\nl
0.20 &6.00 &0.0136 &226 &1.138 &1.6 &1.9 &King\nl
0.25 &1.50 &0.136 &8,318 &1.035 &10.3 &11.0 &King\nl
0.25 &1.75 &0.108 &5,812 &1.049 &8.1 &9.2 &King\nl
0.25 &2.00 &0.0883 &4232 &1.072 &6.2 &7.6 &King\nl
0.25 &3.00 &0.0481 &1,578 &1.104 &3.2 &4.2 &King\nl
0.25 &4.00 &0.0313 &775 &1.154 &2.0 &2.6 &King\nl
0.25 &6.00 &0.0170 &283 &1.175 &1.2 &1.5 &King\nl
0.50 &1.50 &0.272 &15,358 &1.043 &6.9 &10.2 &King\nl
0.50 &1.75 &0.216 &10,986 &1.057 &5.0 &8.4 &King\nl
0.50 &2.00 &0.176 &8,121 &1.079 &3.8 &6.6 &King\nl
0.50 &3.00 &0.0962 &3,108 &1.173 &2.0 &3.2 &King\nl
0.50 &4.00 &0.0625 &1,539 &1.312 &1.4 &2.0 &King\nl
0.50 &6.00 &0.0340 &564 &1.355 &1.0 &1.2 &King\nl
1.10 &1.50 &0.599 &28,529 &1.046 &1.64 &10.2 &King\nl
1.24 &5.0 &0.111 &2,172 &1.60 &0.7 &1.2 &King\nl
0.0042 &2.00 &0.0015 &74 &1.008 &14.5 &14.5 &Plummer\nl
0.0057 &2.00 &0.0020 &101 &1.021 &14.3 &14.3 &Plummer\nl
0.0085 &2.00 &0.0030 &150 &1.021 &14.0 &14.0 &Plummer\nl
0.0101 &2.00 &0.0036 &178 &1.008 &13.9 &13.9 &Plummer\nl
0.0141 &2.00 &0.0050 &249 &1.021 &13.7 &13.7 &Plummer\nl 
0.0286 &2.00 &0.0101 &503 &1.005 &12.8 &12.8 &Plummer\nl
0.113 &2.00 &0.040 &1,960 &1.049 &9.0 &9.3 &Plummer\nl
0.314 &2.00 &0.111 &5,262 &1.066 &5.6 &7.5 &Plummer\nl
2.824 &2.00 &1.000 &33,333 &1.127 &0.014 &7.5 &Plummer\nl
\enddata
\end{deluxetable}

\clearpage

\begin{figure}
\epsscale{0.8}
\plotone{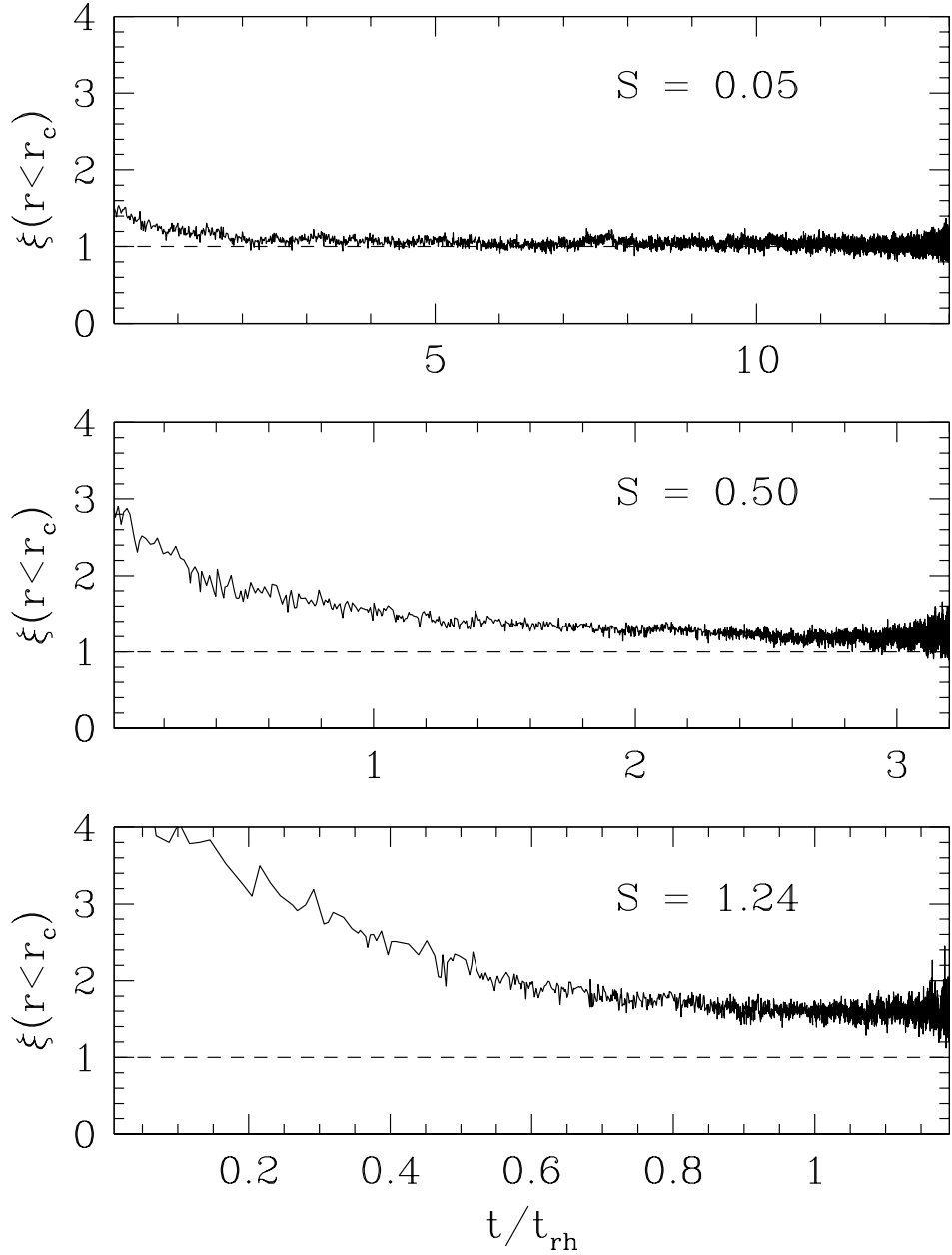}
\caption{Evolution of the temperature ratio in the core for 
$S = 0.05$ and $m_2/m_1 = 1.5$ (top),  $S = 0.5$ and $m_2/m_1 = 3$ (middle), 
and $S = 1.24$ and $m_2/m_1 = 5.0$ (bottom). The minimum temperature ratio 
attained in each calculation increases with $S$.  Time is displayed in units
of the initial half-mass relaxation time ($t_{{\rm rh}}$).  In each case
the evolution is shown until shortly before core collapse.  
Equipartition is clearly not reached prior to core collapse for large $S$.
Notice also that 
core collapse occurs sooner with increasing $S$.  The initial condition in 
each case was a two-component King model with $W_0=6$.
\label{3temprats}}
\end{figure}

\begin{figure}
\epsscale{0.80}
\plotone{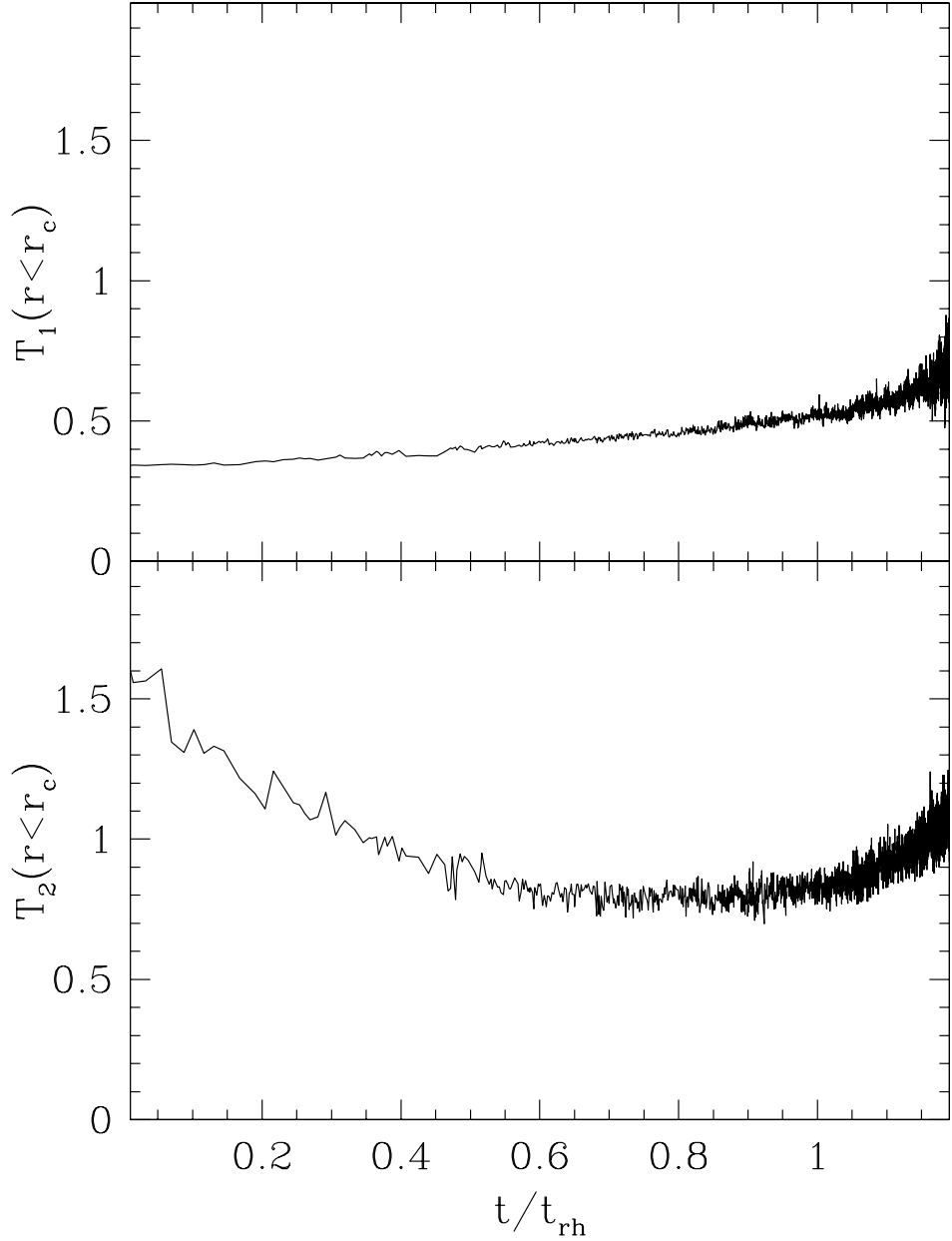}
\caption{Evolution of the core temperature for the lighter stars 
($T_1(r<r_c)$, top) 
and the heavier stars ($T_2(r<r_c)$, bottom) for the case $S = 1.24$ 
and $m_2/m_1 = 
5.0$.  (The ratio of these is shown in the bottom-most plot of 
Fig.~\ref{3temprats}.)  Time is displayed in units of the initial half-mass 
relaxation time ($t_{{\rm rh}}$).
\label{unstable_temps}}
\end{figure}

\begin{figure}
\epsscale{0.8}
\plotone{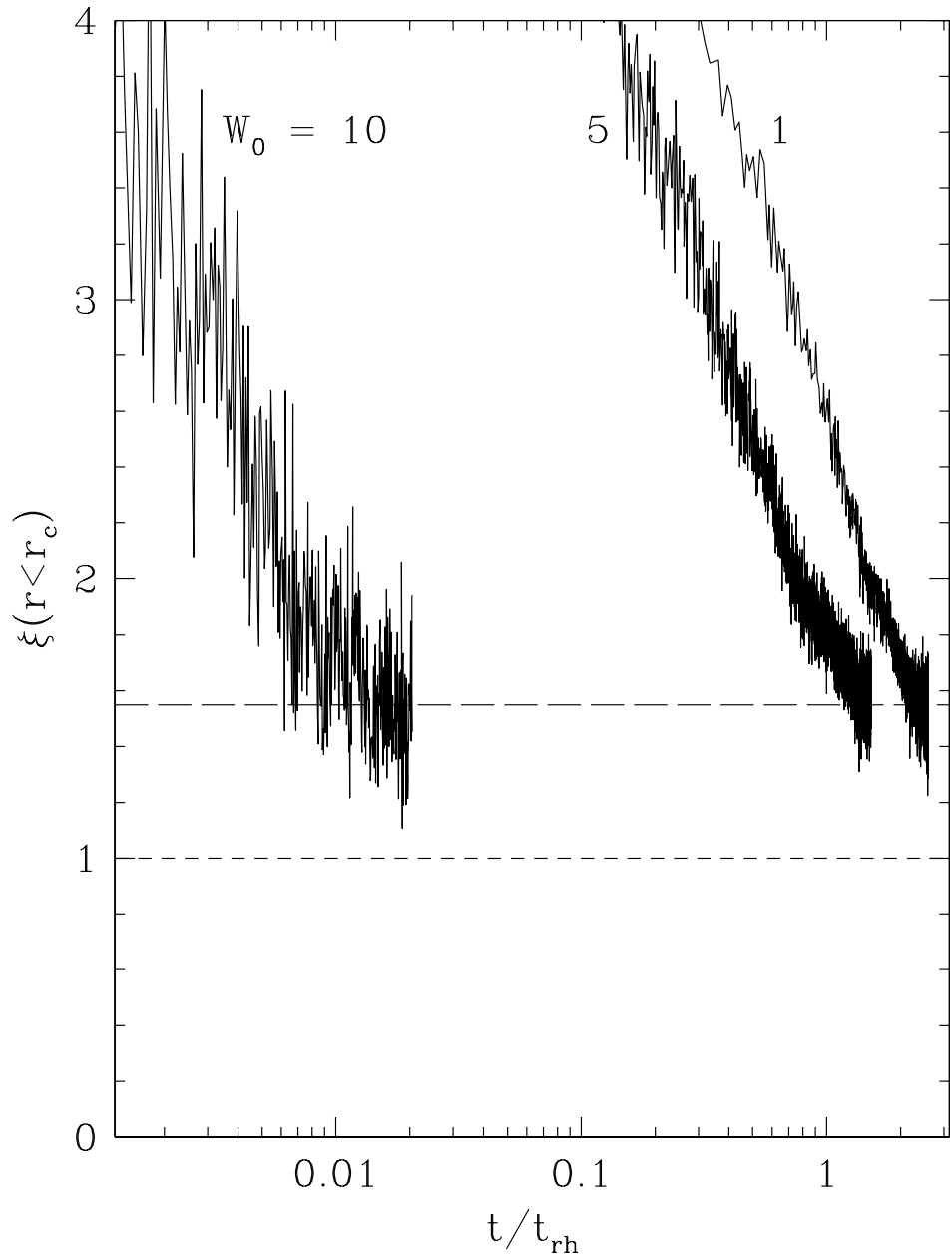}
\caption{Evolution of the core temperature ratio for three calculations with 
different initial values of the dimensionless King model potential $W_0$, but
all with $S = 1$ and $m_2/m_1 = 5$ ($M_2/M_1 \simeq 0.09$).  
From left to right: $W_0 = 10$, $W_0 = 5$, and $W_0 = 1$.  While the relaxation
and core collapse times for these calculations span a wide range, in each case 
the temperature ratio reaches the same minimum value of approximately 1.55.  The 
evolution is shown until shortly before core collapse in each case.  Note that
the logarithmic time scale has compressed the shapes of these curves, so that 
the leveling in the temperature ratio prior to core collapse is not as clearly
apparent as in Fig.~1.
\label{3temprats_3Ws}}
\end{figure}

\begin{figure}
\epsscale{0.8}
\plotone{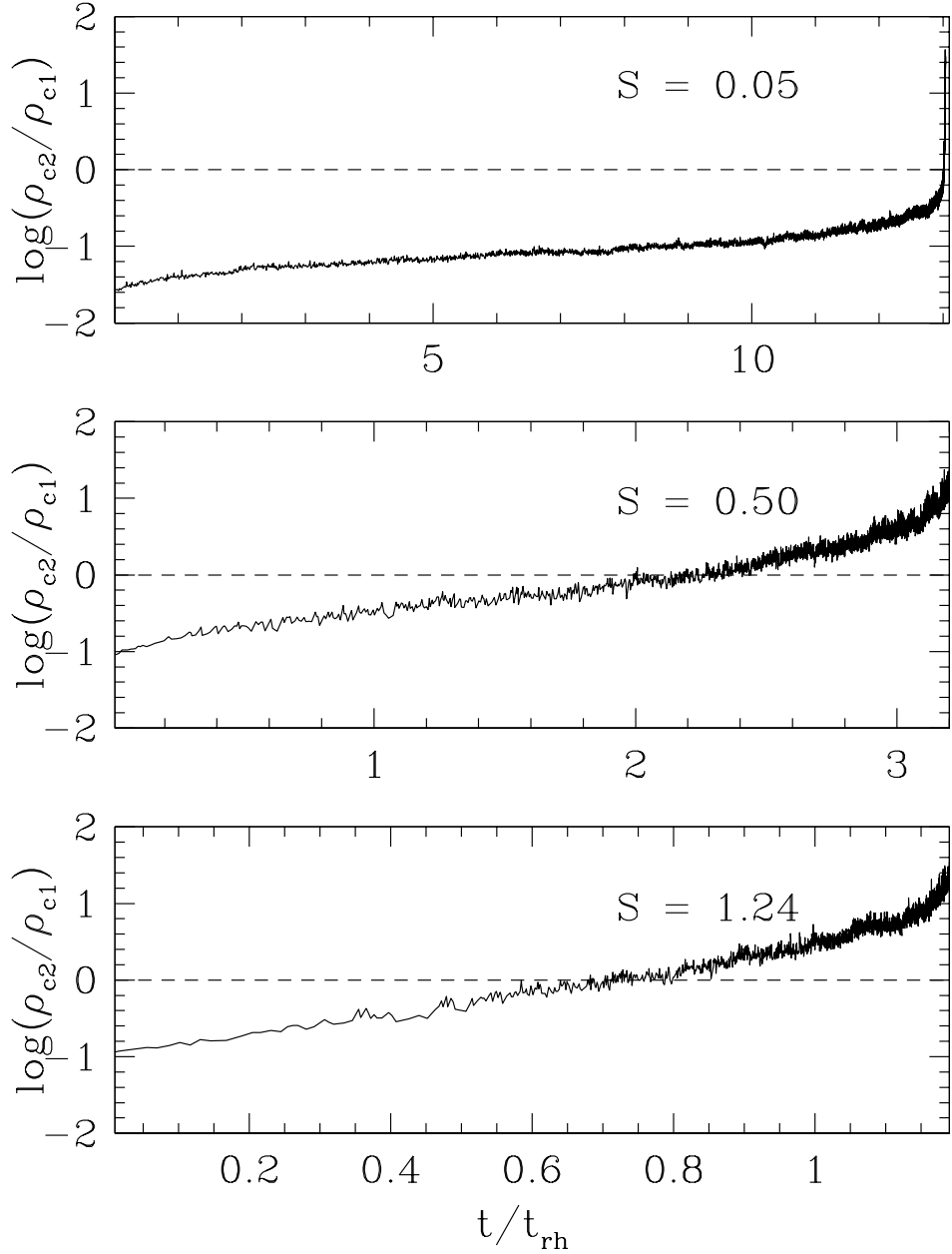}
\caption{Evolution of the mass density ratio in the core 
($\rho_{c2}/\rho_{c1}$) for the same three cases as in Fig.~1:
$S = 0.05$ and $m_2/m_1 = 1.5$ (top),  $S = 0.5$ 
and $m_2/m_1 = 3$ (middle), and $S = 1.24$ and $m_2/m_1 = 5.0$ (bottom). The 
evolution is shown until shortly before core collapse in two cases (middle, 
bottom), and in one case until core collapse (top).  As $S$ increases, the 
time at which equal mass densities are attained in the core ($t_{{\rm \rho}}$) 
occurs sooner with respect to core collapse ($t_{{\rm \rho}} \simeq 
t_{{\rm cc}}$ for the case $S = 0.05$).  
\label{3densrats}}
\end{figure}

\begin{figure}
\epsscale{0.80}
\plotone{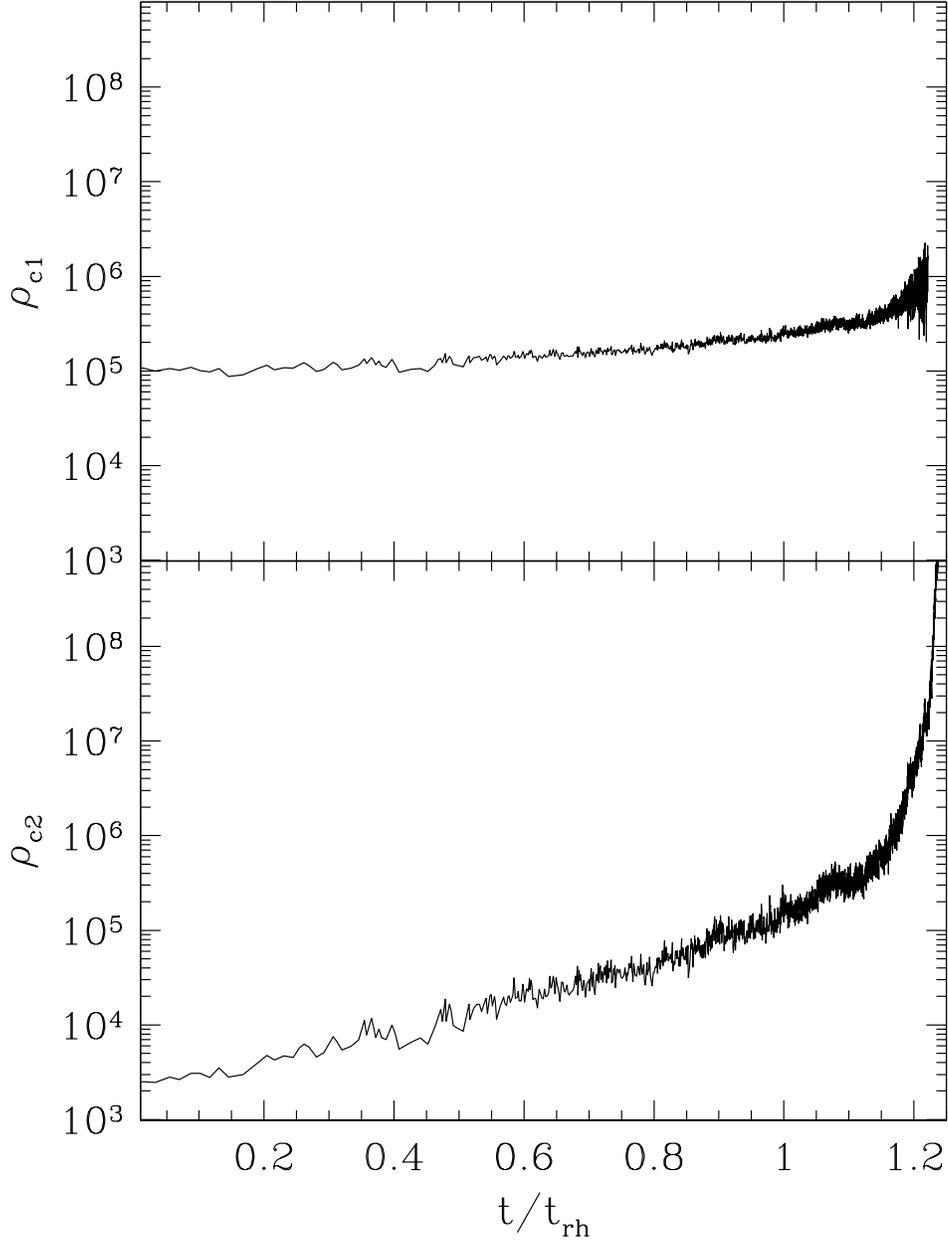}
\caption{Evolution of the core mass density for the lighter stars
(top) and the heavier stars (bottom) for the same case as in Fig.~2:
$S = 1.24$ and $m_2/m_1 = 5.0$.  
(The ratio of these is shown in the bottom-most plot of Fig.~\ref{3densrats}.)  
The evolution of $\rho_{c1}$ is
shown until no lighter stars remain in the core, whereas the evolution of 
$\rho_{c2}$ is shown until core collapse.
\label{unstable_dens}}
\end{figure}

\begin{figure}
\epsscale{1.0}
\plotone{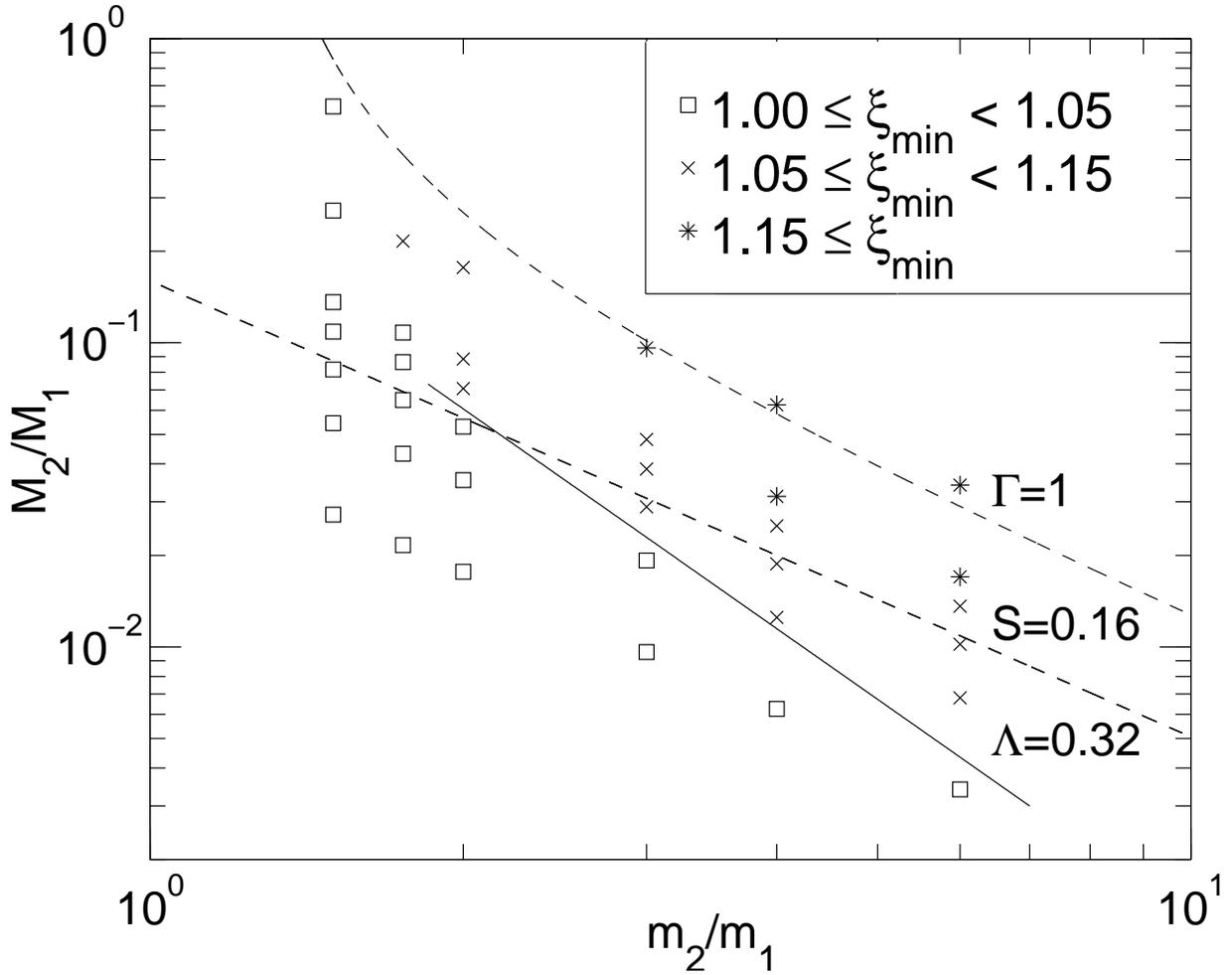}
\caption{Minimum temperature ratio in the core for the 37 calculations in set $A$,
represented here using three symbols at points in the parameter space 
determined by $M_2/M_1$ and $m_2/m_1$.  Also drawn are the Spitzer and 
Lightman-Fall stability boundaries ($S = 0.16$ and $\Gamma = 1$, respectively),
and the boundary $\Lambda = 0.32$ suggested by these results 
(eq.~[\ref{Lambda}]).  Calculations marked with ``$\Box$'' are determined to 
have reached equipartition within our numerical accuracy.
\label{temps}}
\end{figure}

\begin{figure}
\epsscale{1.0}
\plotone{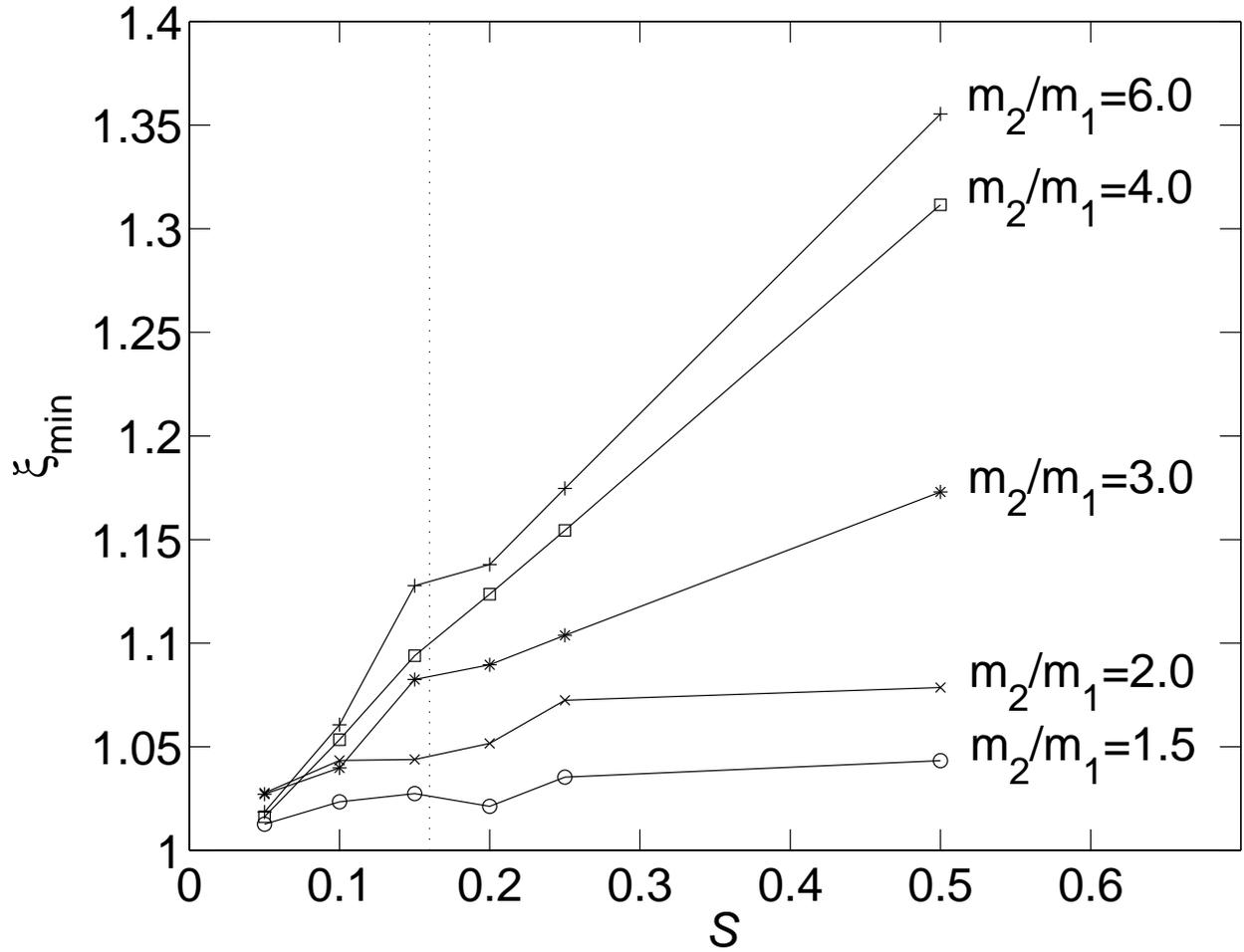}
\caption{Minimum temperature ratio in the core versus $S$ for several values of
$m_2/m_1$.  Also drawn is the Spitzer stability boundary 
($S = 0.16$).
\label{mintemps_S}}
\end{figure}

\begin{figure}
\epsscale{1.0}
\plotone{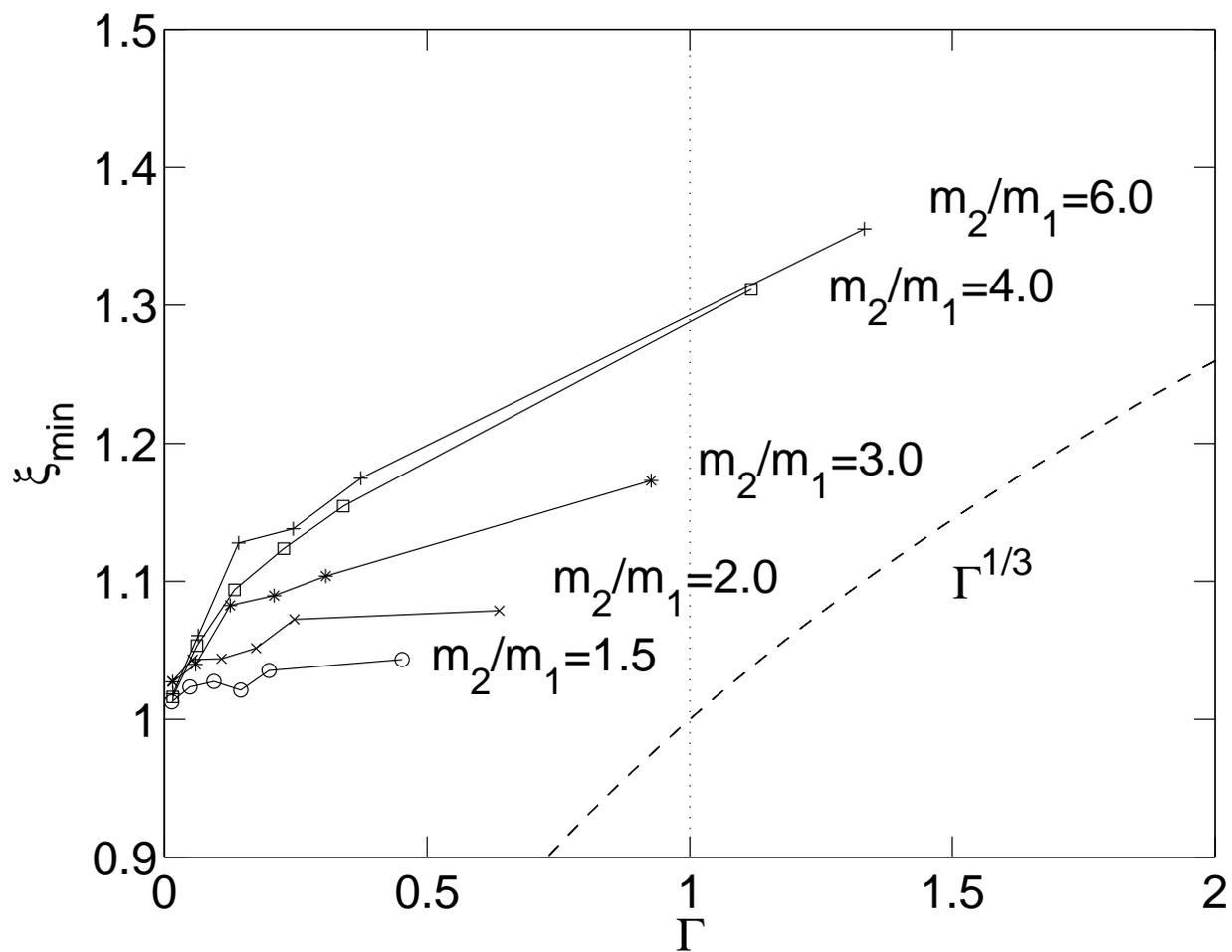}
\caption{Minimum temperature ratio in the core versus $\Gamma$ for several 
values of $m_2/m_1$.  Also drawn is the Lightman-Fall stability boundary 
($\Gamma = 1$) and a theoretical estimate of the minimum core temperature 
ratio, $\Gamma^{1/3}$ (\cite{lightfall78}).
\label{mintemps_Gamma}}
\end{figure}

\begin{figure}
\epsscale{1.0}
\plotone{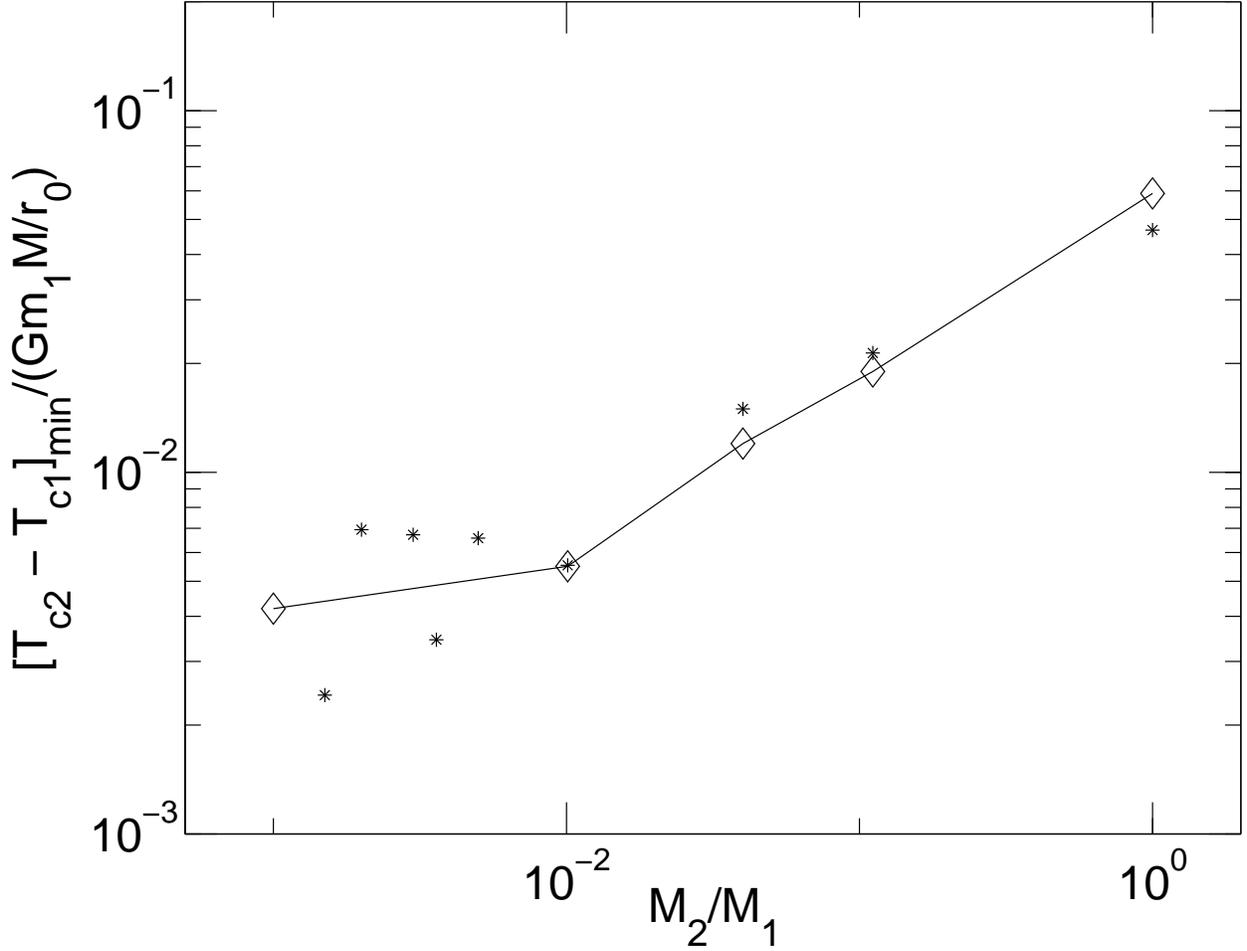}
\caption{Minimum temperature difference in the core versus the total mass ratio
as found in the present study ($\ast$) and by Inagaki \& Wiyanto (1984) 
($\diamond$). Each of these calculations was begun with a two-component 
Plummer model (set $B$) where $m_2/m_1 = 2$.  Here, $r_0$ is the Plummer scale 
length and $M$ is the total cluster mass.  Agreement is very good except in 
the vicinity of small $M_2/M_1$.  In this domain, the temperature of the 
heavier component is calculated using very few stars, so that the 
difference is characterized by a large amount of noise.
\label{compare_plot}}
\end{figure}

\begin{figure}
\epsscale{1.0}
\plotone{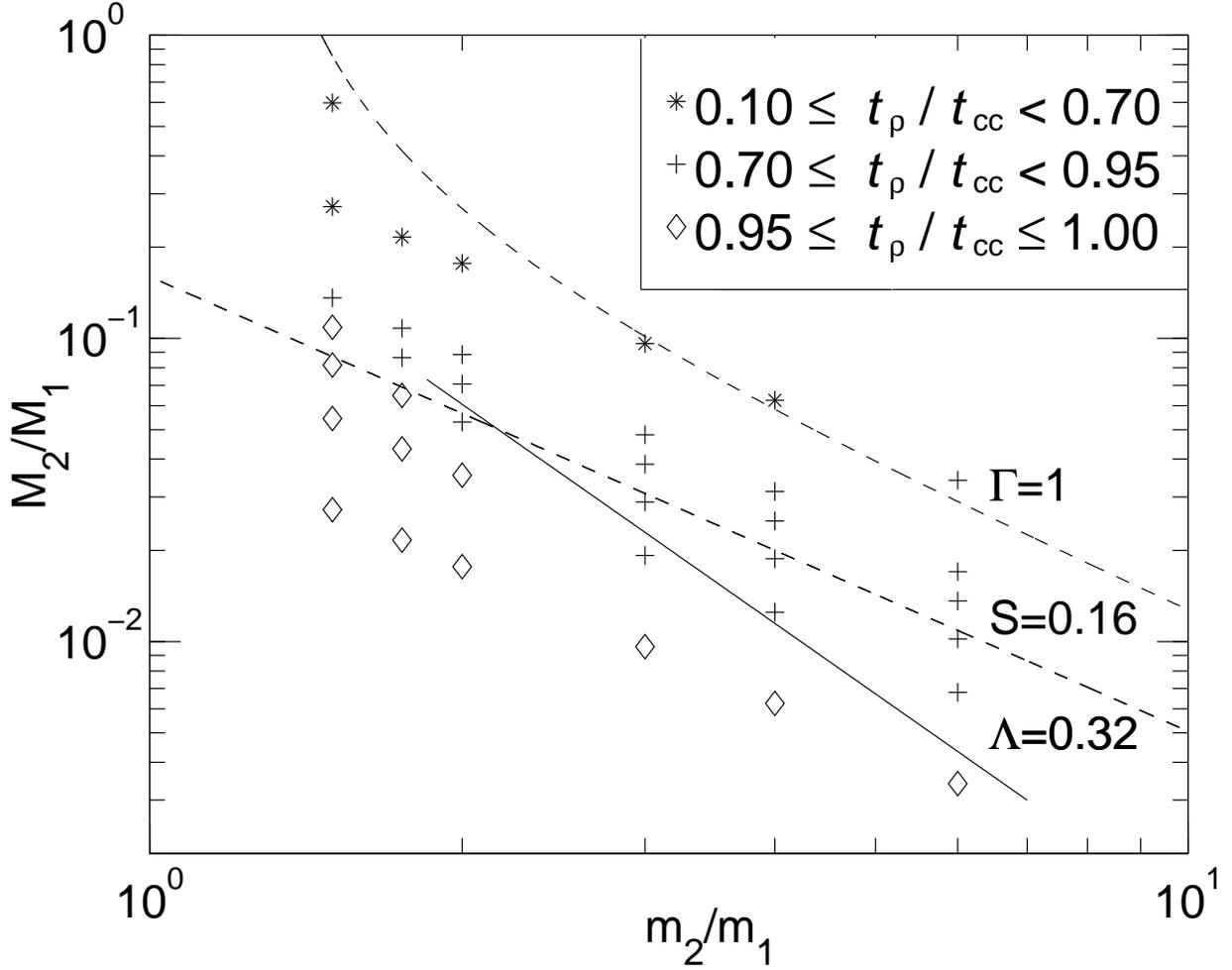}
\caption{Fraction of the core collapse time when equal mass density is attained
in the core ($t_{{\rm \rho}}/t_{{\rm cc}}$) for the 37 calculations in set $A$, 
represented here using three symbols at points in the parameter space 
determined by $M_2/M_1$ and $m_2/m_1$.  Also drawn are the Spitzer and 
Lightman-Fall stability boundaries ($S = 0.16$ and $\Gamma = 1$, respectively),
and the boundary suggested by the results shown in Fig.~\ref{temps} 
($\Lambda=0.32$). Note that our boundary appears to apply approximately
throughout the range $1 < m_2/m_1 < 7$ (in contrast to Fig.~6, where
it does not apply below $m_2/m_1 \simeq 1.75$).
\label{density_tcc}}
\end{figure}

\begin{figure}
\epsscale{1.0}
\plotone{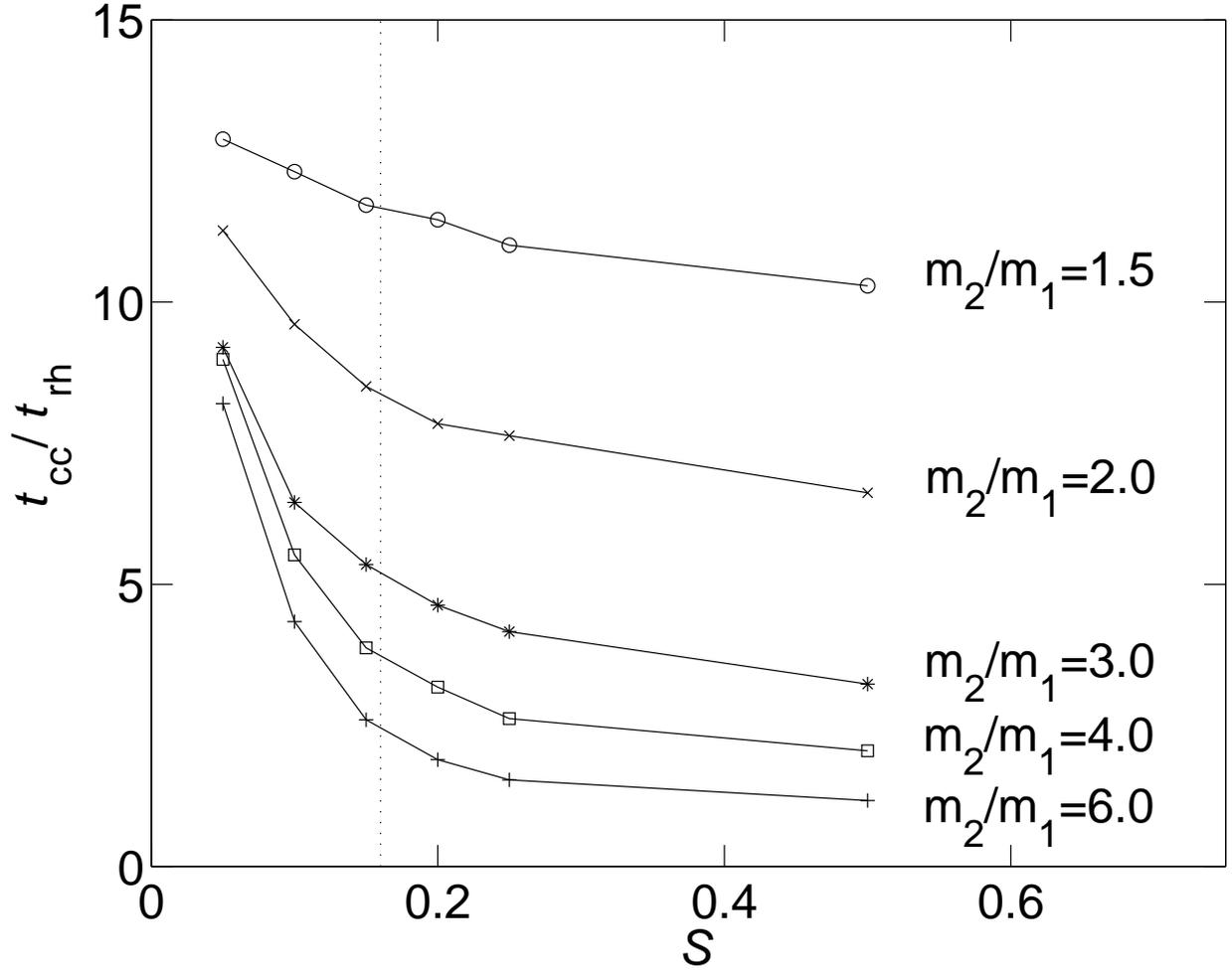}
\caption{Core collapse times versus $S$ for several values of $m_2/m_1$, for 
30 calculations in set $A$.  The initial condition in each case was a 
two-component King model with $W_0=6$.  The times are displayed in units of 
the initial half-mass relaxation time ($t_{{\rm rh}}$).  
\label{tcc_times}}
\end{figure}

\begin{figure}
\epsscale{1.0}
\plotone{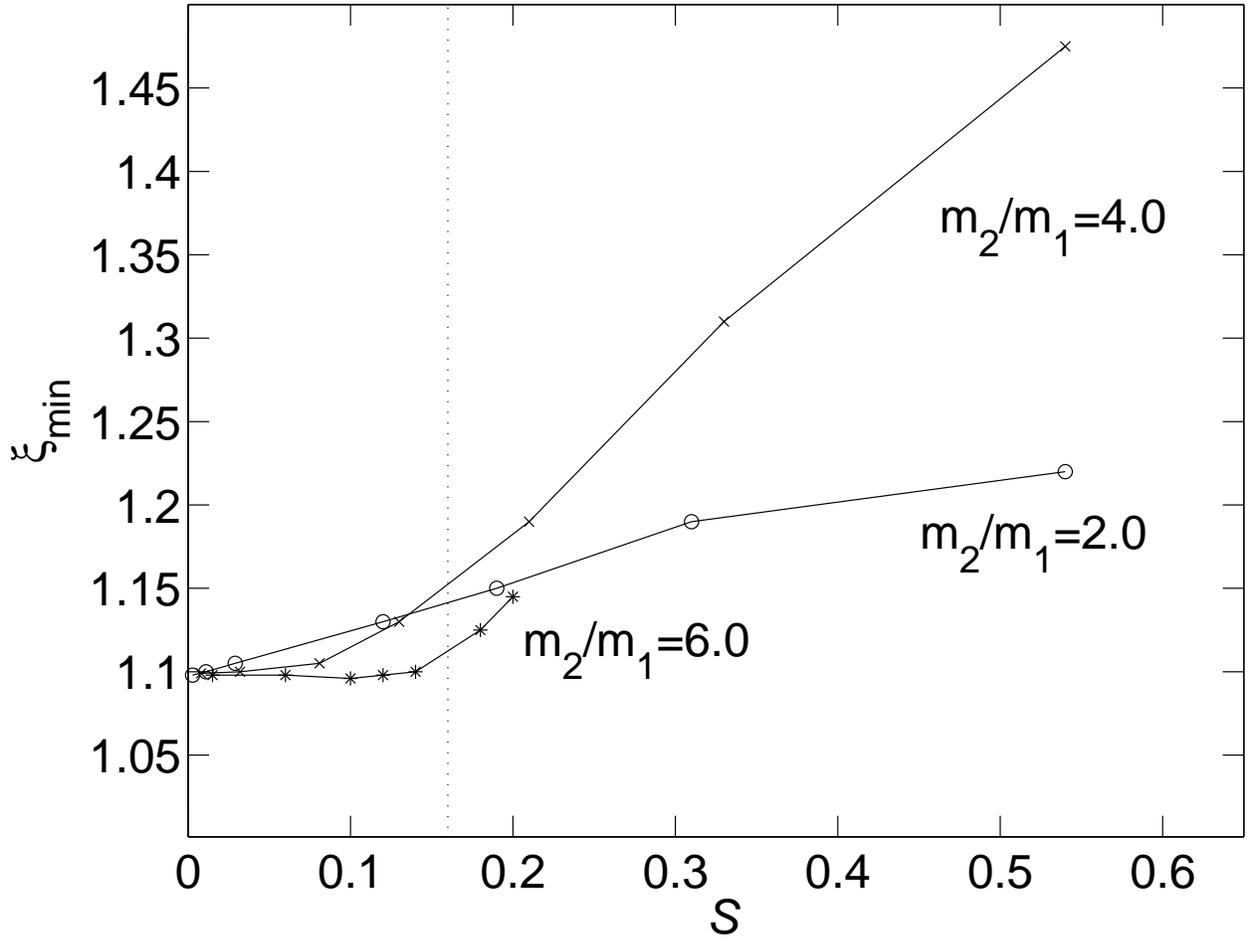}
\caption{Minimum core 
temperature ratio for turning points in a linear series of 
equilibrium models (\cite{katztaff83}).  Also drawn is the Spitzer stability 
boundary ($S = 0.16$).
\label{mintemps_S_katztaff}}
\end{figure}

\begin{figure}
\epsscale{0.80}
\plotone{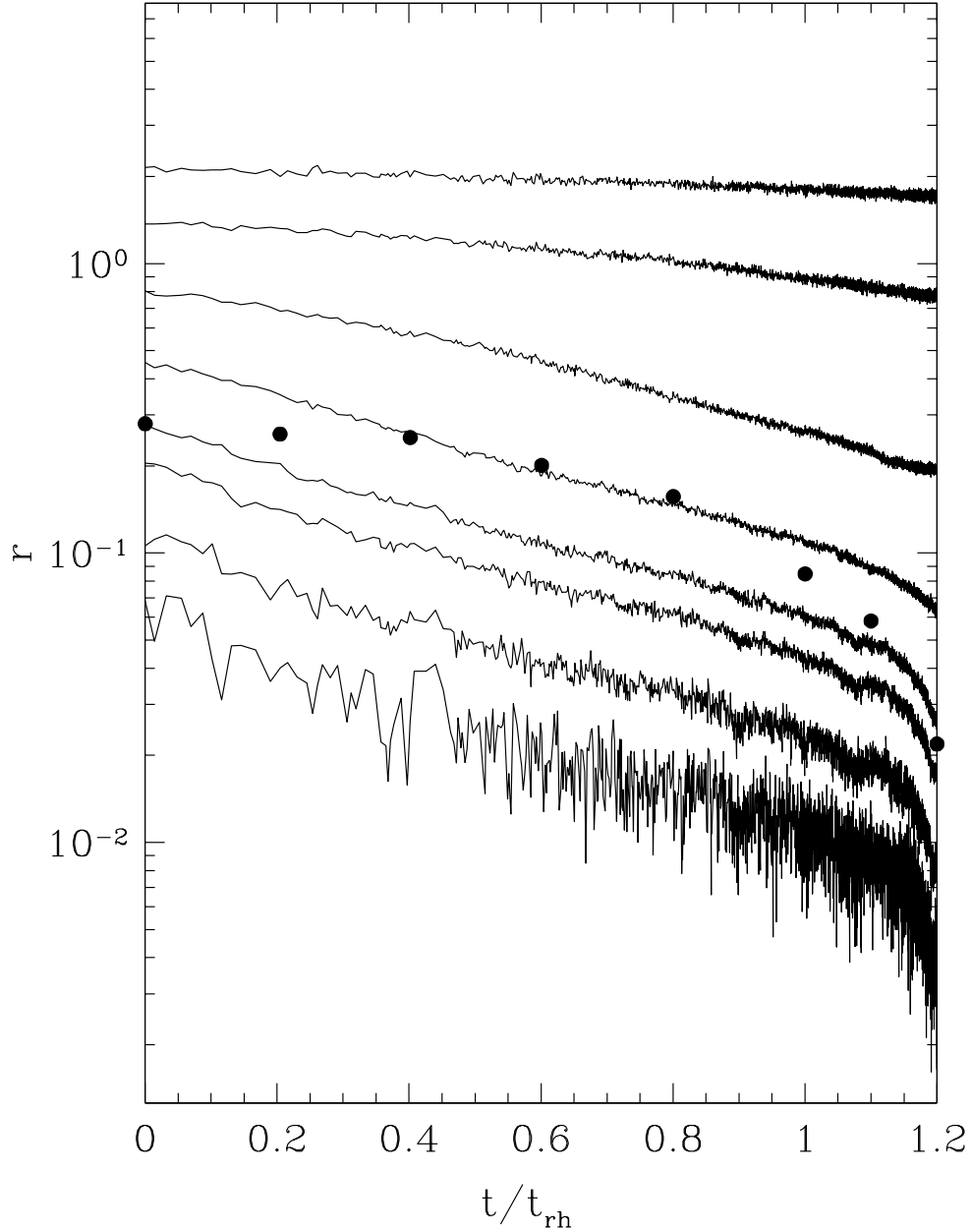}
\caption{Eight Lagrange radii for the heavier component in
the two-component King model with $S=1.24$, $m_2/m_1=5$
(same model as in Fig.~2).  From top to
bottom: the radii containing 90\%, 75\%, 50\%, 25\%, 10\%, 5\%, 1\%, and 0.1\% 
of the total mass in the heavier component.  Also drawn are several points 
in the evolution of the cluster core radius ($\bullet$). Note that 
many stars in the heavier component remain well outside of the core throughout 
the pre-collapse evolution.
\label{S1.241u5extra_lagrange}}
\end{figure}


\clearpage

\begin{thebibliography}{}

\bibitem[Binney \& Tremaine 1989]{bt89}
Binney, J., \& Tremaine, S. 1989, Galactic Dynamics (Princeton, PUP)

\bibitem[Gao et al. 1991]{gao91} 
Gao, B., Goodman, J., Cohn H., \& Murphy, B.  1991, \apj, 370, 567

\bibitem[H\'enon 1971]{hen71} 
H\'enon, M.  1971, \apss, 14, 151

\bibitem[Inagaki 1985]{inag85}
Inagaki, S. 1985, in IAU Symposium 113, The Dynamics of Star Clusters,
ed. J.~Goodman \& P.~Hut (Dordrecht: Reidel), 269 

\bibitem[Inagaki \& Wiyanto 1984]{inagwiy84} 
Inagaki, S., \& Wiyanto, P.  1984, \pasj, 36, 391

\bibitem[Joshi, Rasio, \& Portegies Zwart 2000]{joshraszwart00} 
Joshi, K.~J., Rasio, F.~A., \& Portegies Zwart, S. 2000a, submitted to \apj, [astro-ph/9909115]

\bibitem[Joshi, Nave, \& Rasio 2000]{joshnaveras00} 
Joshi, K.~J., Nave, C.~P., \& Rasio, F.~A. 2000b, submitted to \apj, [astro-ph/9912155]

\bibitem[Katz \& Taff 1983]{katztaff83} 
Katz, J., \& Taff, L.~G.  1983, \apj, 264, 476  

\bibitem[Kim, Lee, \& Goodman 1998]{kimleegood98} 
Kim, S.~S., Lee, H.~M., \& Goodman, J.  1998, \apj, 495, 786

\bibitem[King 1966]{king66} 
King, I.~R., 1966, \aj, 71, 64

\bibitem[Kondrat'ev \& Ozernoy 1982]{kondroz82} 
Kondrat'ev, B.~P., \& Ozernoy, L.~M.  1982, \apss, 84, 431

\bibitem[Kulkarni et al. 1993]{kulk93}
Kulkarni, S.R., Hut, P., \& McMillan, S.L.W. 1993, Nature, 364, 421

\bibitem[Lightman \& Fall 1978]{lightfall78} 
Lightman, A.~P., \& Fall, S.~M. 1978, \apj, 221, 567

\bibitem[Merritt 1981]{merr81} 
Merritt, D.  1981, \aj, 86, 318

\bibitem[Meylan \& Heggie 1997]{meyheg97}
Meylan, G., \& Heggie, D.C. 1997, A\&AR, 8, 1

\bibitem[Murphy et al. 1998]{murph98}
Murphy, B.W., Moore, C.A., Trotter, T.E., Cohn, H.N., \& Lugger, P.M. 1998,
AAS Meeting 193, 60.01

\bibitem[Portegies Zwart \& McMillan 2000]{zwartmcmill00} 
Portegies Zwart, S., \& McMillan, S. 2000, ApJ Letters, in press, [astro-ph/9910061]

\bibitem[Quinlan 1996]{quin96} 
Quinlan, G.~D.,  1996, New Astronomy, 1, 255 

\bibitem[Sigurdsson \& Hernquist 1993]{sighern93}
Sigurdsson, S., \& Hernquist, L. 1993, Nature, 364, 423

\bibitem[Spitzer 1969]{spitz69} 
Spitzer, L., Jr.  1969, \apj, 158, L139

\bibitem[Spitzer 1987]{spitz87} 
Spitzer, L., Jr. 1987, Dynamical Evolution of Globular Clusters 
(Princeton University Press)

\bibitem[Spitzer \& Hart 1971]{spithart71} 
Spitzer, L., Jr., \& Hart, M.~H.  1971, \apj, 166, 483

\bibitem[Spitzer \& Shull 1975]{spitshull75} 
Spitzer, L., Jr., \& Shull, J.M. 1975, \apj, 201, 773

\bibitem[Yoshizawa et al. 1978]{yosh78} 
Yoshizawa, M., Inagaki, S., Nishida, M.~T., Kato, S., Tanaka, Y., 
\& Watanabe, Y.  1978, \pasj, 30, 279

 
\end{thebibliography}
\end{document}